\documentclass[twocolumn,aps,longbibliography]{revtex4-2}

\usepackage{graphicx}
\usepackage{colordvi}
\usepackage{color}

\definecolor{xred}{rgb}{1,0,0}

\begin{document}



\title{Unavoidability of nonclassicality loss in PT-symmetric systems}

\author{Jan Pe\v{r}ina Jr.}
\email{jan.perina.jr@upol.cz} \affiliation{Joint Laboratory of
Optics, Faculty of Science, Palack\'{y} University, Czech
Republic, 17. listopadu 12, 771~46 Olomouc, Czech Republic}

\author{Adam Miranowicz}
\affiliation{Institute of Spintronics and Quantum Information,
Faculty of Physics, Adam Mickiewicz University, 61-614 Pozna\' n,
Poland}

\author{Joanna K. Kalaga}
\affiliation{Quantum Optics and Engineering Division, Faculty of
Physics and Astronomy, University of Zielona G\' ora, Prof. Z.
Szafrana 4a, 65-516 Zielona G\' ora, Poland}

\author{Wies\l aw Leo\' nski}
\affiliation{Quantum Optics and Engineering Division, Faculty of
Physics and Astronomy, University of Zielona G\' ora, Prof. Z.
Szafrana 4a, 65-516 Zielona G\' ora, Poland}

\date{\today}

\begin{abstract}
We show that the loss of nonclassicality (including quantum
entanglement) cannot be compensated by the (incoherent)
amplification of $\mathcal{PT}$-symmetric systems. We address this
problem by manipulating the quantum fluctuating forces in the
Heisenberg-Langevin approach. Specifically, we analyze the
dynamics of two nonlinearly coupled oscillator modes in a
$\mathcal{PT}$-symmetric system. An analytical solution allows us
to separate the contribution of reservoir fluctuations from the
evolution of quantum statistical properties of the modes.
In general, as reservoir fluctuations act constantly, the complete
loss of nonclassicality and entanglement is observed for long
times. To elucidate the role of reservoir fluctuations in a
long-time evolution of nonclassicality and entanglement, we
consider and compare the predictions from two alternative models
in which no fatal long-time detrimental effects on the
nonclassicality and entanglement are observed. This is so as, in
the first semiclassical model, no reservoir fluctuations are
considered at all. This, however, violates the
fluctuation-dissipation theorem. The second, more elaborated,
model obeys the fluctuation-dissipation relations as it partly
involves reservoir fluctuations. However, to prevent from the
above long-time detrimental effects, the reservoir fluctuations
have to be endowed with the nonphysical properties of a sink
model. In both models, additional incorporation of the omitted
reservoir fluctuations results in their physically consistent
behavior. This behavior, however, predicts the gradual loss of the
nonclassicality and entanglement. Thus the effects of reservoir
fluctuations related to damping cannot be compensated by those
related to amplification. This qualitatively differs from the
influence of damping and amplification to a direct coherent
dynamics of $\mathcal{PT}$-symmetric systems in which their mutual
interference results in a periodic behavior allowing for
nonclassicality and entanglement at arbitrary times.
\end{abstract}

\maketitle

\section{Introduction}

Open systems can be found in many areas of physics, chemistry and
biology. Their rigorous description is based upon using the
(generalized) master equations that is, however, demanding. In
special cases, in which damping and amplification in the analyzed
system are in balance, their description via an appropriate
non-Hermitian parity-time ($\mathcal{PT}$) symmetric Hamiltonian
represents an attractive alternative. This is possible due to the
fact that such Hamiltonians, though being non-Hermitian, are
endowed with real spectra. Non-Hermitian ($\mathcal{PT}$)
symmetric Hamiltonians have become attractive owing to the works
by Bender et al. ~\cite{Bender1998,Bender1999,Bender2003}. The
presence of exceptional points (EPs) is another important feature
of such Hamiltonians. At EPs, that are, in a certain sense,
singular points in parameter spaces, the systems exhibit special
properties and physical effects (for details, see
reviews~\cite{Ozdemir2019,Miri2019}). They may be used, e.g., for
enhanced sensing~\cite{Liu2016,Chen2017,Hodaei2017}, enhanced
nonlinear
interactions~\cite{He2015,Vashahri2017,PerinaJr2019x,PerinaJr2019y},
unidirection light propagation~\cite{Peng2014,Chang2014}, and
invisibility~\cite{Lin2011,Regen2012}.

For such reasons and concentrating on optics, numerous classical
and semiclassical $\mathcal{PT}$-symmetric systems were analyzed
in the areas of optical waveguides~\cite{Turitsyna2017,Xu2018},
optical coupled
structures~\cite{El-Ganainy2007,Ramezani2010,Zyablovsky2014,Ogren2017},
coupled optical
microresonators~\cite{Peng2014,Peng2014a,Liu2016,Zhou2016,Arkhipov2019,
Minganti2020,Minganti2021,Minganti2021a}, optical
lattices~\cite{Graefe2011,Miri2012,Ornigotti2014,Shui2019} or even
chaotic systems~\cite{Szewczyk2022}. Models based on
$\mathcal{PT}$-symmetric Hamiltonians and their EPs can also be
found in microwave photonics~\cite{Quijandria2018},
plasmonics~\cite{Benisty2011} (for a review see~\cite{Tame2013}),
electronics~\cite{Schindler2011,Chen2019},
metamaterials~\cite{Kang2013}, cavity
optomechanics~\cite{Jing2014,Harris2016,Jing2017}, and
acoustics~\cite{Zhu2014,Alu2015}. Moreover, problems related to
$\mathcal{PT}$ symmetry were considered in the context of quantum
steering~\cite{Duc2021}, stability of the hydrogen
molecule~\cite{Wrona2020}, and even finding energy levels for
hydrogen bridge in nanojunctions with metallic
anchors~\cite{Domagalska2022}.

Whereas $\mathcal{PT}$-symmetric non-Hermitian Hamiltonians have
been extraordinarily successful in describing numerous effects in
classical and semiclassical systems, their application to fully
quantum systems is not straightforward. Damping and amplification
that are indispensable parts of such Hamiltonians cause
back-action of system's surroundings (reservoir) that influences
the dynamics of the analyzed system itself. The strength of
back-action is proportional to the level of damping and
amplification (according to the fluctuation-dissipation theorem
\cite{Scheel2018}). The back-action that is typically described by
(random) reservoir fluctuating forces disturbs quantum coherence
in a given system. This results in the gradual loss of
nonclassicality and entanglement (quantum correlations) of the
system states. This is rather limiting, e.g., for quantum
nonlinear optics, in which the generation of nonclassical and
entangled states has been extensively studied
\cite{He2015,Vashahri2017,PerinaJr2019x,PerinaJr2019y}.

We note that, in Refs.~\cite{Ju2019,Ju2022}, an alternative
description of a quantum $\mathcal{PT}$-symmetric system was
suggested using an equation of motion derived for the metric of
the Hilbert space induced by the system. For any Hermitian system,
such a metric is trivially equal to one, but it can be highly
nontrivial for non-Hermitian systems. However, by neglecting a
proper Hilbert-space metric in describing the evolution of systems
with non-Hermitian Hamiltonians, one can seemingly violate the
basic no-go theorems in quantum mechanics, including those in
quantum information, as explicitly demonstrated in \cite{Ju2019}.
Alternatively, when analyzing non-Hermitian quantum systems
quantum jumps can be included to follow consistent quantum
evolution of such systems, as shown in Refs.~\cite{Minganti2019,
Minganti2020} in the context of EPs. In general, there is a
question to which extent $\mathcal{PT}$-symmetric non-Hermitian
Hamiltonians provide a suitable tool for describing more complex
physical systems \cite{Purkayastha2020}.

The question arises whether the action of reservoir fluctuations
has to inevitably result in the complete loss of nonclassicality
and entanglement in a long-time evolution of quantum systems. To
answer this question, we analyze here the role of different types
of reservoir fluctuating forces (with different
fluctuation-dissipation relations) in the dynamics of nonclassical
properties in a system of two coupled oscillator modes with one
mode damped and the other amplified. First, we consider a quantum
statistical model in which both modes interact with proper
physical reservoirs whose elimination from the description of a
master system results in the quantum Heisenberg-Langevin
equations. Their solution describes the above-discussed loss of
nonclassicality and entanglement for long times. To understand the
role of reservoir fluctuations in the evolution of the
nonclassicality and entanglement, we consider the corresponding
semiclassical model, in which no reservoir fluctuations are
considered to compensate for damping and amplification in the
master system. This, on one side, leads to a periodic solution
that allows for the long-time nonclassicality and entanglement,
but, on the other side, it violates the fluctuation-dissipation
theorem and, thus, disturbs quantum consistency of the model. To
keep quantum consistency, we formulate another model that
partially involves the reservoir fluctuating forces such that the
fluctuation-dissipation relations are satisfied. This leads,
similarly as in the semiclassical model, to a periodic solution
that admits the long-time nonclassicality and entanglement. The
revealed ideal reservoir is common for both oscillator modes.
Moreover, its properties resemble those of the sink models
\cite{Silinsh1994} that remove energy (particles) from the master
system. Such reservoir properties are considered as nonphysical.
We note that when the missing parts of the reservoir fluctuating
forces in both models are taken into account, the system evolution
looses its periodicity together with the long-time nonclassicality
and entanglement.

Detailed analysis of both models with partially suppressed
reservoir fluctuations leads us to the general conclusion: When
usual physical reservoirs with classical properties are considered
to compensate for damping and amplification in the master system,
the gradual loss of system's nonclassicality and entanglement in
its evolution has to inevitably occur as a consequence of the
action of reservoir fluctuations.

This means, among others, that the analysis of quantum systems
based on the $\mathcal{PT}$-symmetric non-Hermitian Hamiltonians
is principally limited to shorter times.

The paper is organized as follows. In Sec.~II, the model of two
coupled oscillator modes is presented and its dynamics is
completely solved including the reservoir contribution. In
Sec.~III, the properties of an ideal reservoir that do not destroy
the long-time nonclassicality and entanglement are derived. The
nonclassicality and entanglement for the Gaussian states in the
suggested quantum models and in the semiclassical model with no
reservoir fluctuations are analyzed in Sec.~IV. The predictions of
the models for specific time are compared in Sec.~V. Conclusions
are drawn in Sec.~VI.

\section{Model of two coupled oscillator modes and their
evolution}

By introducing the photon annihilation ($ \hat{a}_j $) and
creation ($ \hat{a}_j^\dagger $) operators of the considered
oscillator modes labelled as 1 and 2, we can write the appropriate
interaction Hamiltonian $ \hat{H} $ of the system as
follows~\cite{Perina1991}:
\begin{eqnarray}  
 \hat{H} = \left[ \epsilon \hat{a}_1^\dagger\hat{a}_2 +     
   \kappa \hat{a}_1\hat{a}_2 + {\rm h.c.} \right] 
   + \left[  \hat{a}_1 \hat{l}_1^\dagger + \hat{a}_2 \hat{l}_2^\dagger
   + {\rm h.c.} \right] ,
\label{1}
\end{eqnarray}
where $ \epsilon $ describes linear exchange of energy (photons)
between modes 1 and 2. The coupling constant $ \kappa $ originates
in parametric down-conversion \cite{Mandel1995} that creates and
annihilates photons in modes 1 and 2 in pairs and, thus, is
responsible for the generation of nonclassical states in the
system. Symbol h.c. replaces the Hermitian conjugated terms. We
assume that mode 1 is damped with a damping constant $ \gamma $
and mode 2 is amplified with the same amplification constant $
\gamma $ ($\mathcal{PT}$-symmetry). The annihilation ($ \hat{l}_j
$) and creation ($ \hat{l}_j^\dagger $) operators of the
corresponding Langevin fluctuating operator forces describe the
reservoir back-action to the damping and amplification.

To guarantee the quantum consistency of the system evolution, the
Langevin fluctuating operator forces are usually modelled by two
independent quantum random Gaussian processes with the following
correlation functions~\cite{Meystre2007,Agarwal2012,Perinova2019}:
\begin{eqnarray} 
 \langle\hat{l}_1(t)\rangle = \langle\hat{l}_1^\dagger(t)\rangle = 0,  \hspace{3mm}
 \langle\hat{l}_2(t)\rangle = \langle\hat{l}_2^\dagger(t)\rangle = 0,\nonumber \\
 \langle\hat{l}_1^\dagger(t)\hat{l}_1(t')\rangle = 0,
  \hspace{3mm} \langle\hat{l}_1(t)\hat{l}_1^\dagger(t')\rangle =
  2\gamma \delta(t-t'), \nonumber \\
 \langle\hat{l}_2^\dagger(t)\hat{l}_2(t')\rangle = 2\gamma \delta(t-t'),
  \hspace{3mm} \langle\hat{l}_2(t)\hat{l}_2^\dagger(t')\rangle =0;
\label{2}
\end{eqnarray}
the remaining second-order correlation functions are zero. Symbol
$ \delta $ stands for the Dirac function. Whereas the Langevin
forces of mode 1 correspond to the reservoir two-level atoms in
the ground state, the Langevin forces of mode 2 arise for the
excited reservoir two-level atoms. We note that $
[\hat{a}_j,\hat{a}_j^\dagger] = 1 $ for $ j=1,2 $ are the only
nonzero commutation relations among the operators $ \hat{a}_j $
and $ \hat{a}^\dagger_j $.

The Heisenberg equations derived from the Hamiltonian $ \hat{H} $
in Eq.~(\ref{1}) can conveniently be written in the matrix form
\begin{eqnarray} 
 \frac{d\hat{\bf A}(t)}{dt} &=& {\bf M} \hat{\bf A}(t)
  + \hat{\bf L}(t),
\label{3}  \\
 & & {\bf M} =   \left[ \begin{array}{cccc}
   -\gamma & 0 & \epsilon & \kappa \\
   0 & -\gamma & -\kappa & -\epsilon \\
   \epsilon & \kappa & \gamma & 0 \\
   -\kappa & -\epsilon & 0 & \gamma \end{array} \right]
\label{4}
\end{eqnarray}
assuming real $ \epsilon $ and $ \kappa $ and using the vectors $
\hat{\bf A}^{\rm T} = (\hat{a}_1,\hat{a}_1^\dagger,
\hat{a}_2,\hat{a}_2^\dagger) $ and $ \hat{\bf L}^{\rm T} =
(\hat{l}_1,\hat{l}_1^\dagger, \hat{l}_2,\hat{l}_2^\dagger) $. We
note that the positions of EPs of the system described by the
Heisenberg equations~(\ref{3}) including their degeneracies were
discussed in \cite{PerinaJr2022} from the point of view of the
Liouvillian EPs. We also note that the model described by the
Heisenberg equations (\ref{3}) can equivalently be formulated
using a master equation (for details, see, e.g.,
Ref.~\cite{Vogel2001}). In this case, details about the inclusion
of the reservoirs described in Eq.~(\ref{2}) can be found, e.g.,
in Ref.~\cite{PerinaJr2022}.

The solution of the linear operator equations in (\ref{3}) can be
expressed using the evolution matrix $ \hat{P}(t,t')
$~\cite{PerinaJr2000}:
\begin{eqnarray}  
 \hat{\bf A}(t) &=& {\bf P}(t,0)\hat{\bf A}(0)
    + \hat{\bf F}(t),
\label{5}   \\
  \hat{\bf F}(t) &=& \int_{0}^t dt' {\bf P}(t,t')
  \hat{\bf L}(t').
\label{6}
\end{eqnarray}
The evolution matrix $ {\bf P}(t,t') $ arises as a solution of the
equation
\begin{equation} 
 \frac{d{\bf P}(t,t')}{dt} = {\bf M}{\bf P}(t,t')
\label{7}
\end{equation}
with the boundary condition $ {\bf P}(t,t') $ equal to the unity
matrix. The solution is written as:
\begin{equation} 
 {\bf P}(t,t') = \exp[ {\bf M} (t-t')].
\label{8}
\end{equation}
Equation (\ref{6}) for the fluctuating forces $ \hat{\bf F} $
leads to the correlation functions as follows~\cite{PerinaJr2000}:
\begin{eqnarray} 
 \langle \hat{\bf F}(t)\rangle &=& \int_{0}^t dt' {\bf P}(t,t')
  \langle \hat{\bf L}(t') \rangle, \nonumber \\
  \langle \hat{\bf F}(t)^{} \hat{\bf F}^{\dagger {\rm T}}(t)\rangle &=& \int_{0}^t d\tilde{t}
  \int_{0}^t d\tilde{t}' {\bf P}(t,\tilde{t}) \langle \hat{\bf L}(\tilde{t})
  \hat{\bf L}^{\dagger {\rm T}}(\tilde{t}')\rangle {\bf P}^{\dagger {\bf T}}(t,\tilde{t}'). \nonumber \\
 & &
\label{9}
\end{eqnarray}

Once the diagonal form of the dynamical matrix $ {\bf M} $ in
Eq.~(\ref{4}) is revealed the solution of the model can be
expressed analytically. Relying on the block structure of the
matrix $ {\bf M} $ we find the following result:
\begin{eqnarray} 
 {\bf M} &=& {\bf T} {\Lambda}_{\bf M} {\bf T}^{-1};
\label{10} \\
 & & {\Lambda}_{\bf M} = \mu \;  {\rm diag}(1,1,-1,-1) ,
\label{11}  \\
 & & {\bf T} = ({\bf T}_1,{\bf T}_2,{\bf T}_3,{\bf T}_4),
\label{12} \\
 & & {\bf T}_{1,2} = \frac{1}{2\sqrt{\epsilon}} \Bigl(
  \zeta^\pm, -\zeta^\mp, \pm \zeta^\pm \psi^+, \mp \zeta^\mp \psi^+
   \Bigr), \nonumber \\
 & & {\bf T}_{3,4} = \frac{1}{2\sqrt{\epsilon}} \Bigl(
  \zeta^\pm, -\zeta^\mp, \mp \zeta^\pm \psi^-, \pm \zeta^\mp \psi^-
   \Bigr), \nonumber \\
 & & {\bf T}^{-1} = ({\bf T}^{-1}_1,{\bf T}^{-1}_2,{\bf T}^{-1}_3,{\bf T}^{-1}_4),
\label{13} \\
 & & {\bf T}^{-1}_{1,2} = \frac{\sqrt{\epsilon}}{2\sqrt{\mu}} \Bigl(
  \zeta^\pm \psi^-, - \zeta^\mp \psi^- ,
  \zeta^\pm \psi^+, - \zeta^\mp \psi^+   \Bigr), \nonumber \\
 & & {\bf T}^{-1}_{3,4} = \frac{\sqrt{\epsilon}}{2\sqrt{\mu}} \left(
  \zeta^\pm, \zeta^\mp, -\zeta^\pm, -\zeta^\mp  \right), \nonumber
\end{eqnarray}
and $ \xi = \sqrt{\epsilon^2-\kappa^2} $, $ \zeta^\pm =
\sqrt{\epsilon \pm \xi} $, $ \mu =
\sqrt{\epsilon^2-\kappa^2-\gamma^2} $, and $ \psi^\pm = (\mu \pm
i\gamma) /\xi $.

By determining the evolution matrix $ {\bf P} $ in Eq.~(\ref{8})
with the help of the decomposition of the dynamical matrix $ {\bf
M} $ in Eq.~(\ref{10}), we can express the solution of the
Heisenberg-Langevin equations in Eq.~(\ref{5}) in the form
\begin{equation}  
 \hat{\bf a}(t) = {\bf U}(t)\hat{\bf a}(0)
  + {\bf V}(t) \hat{\bf a}^\dagger(0) + \hat{\bf f}(t)
\label{14}
\end{equation}
using the definitions $ \hat{\bf a}^{\rm T} \equiv
(\hat{a}_1,\hat{a}_2) $, $ U_{j,k}(t) = P_{2j-1,2k-1}(t,0) $, $
V_{jk}(t) = P_{2j-1,2k}(t,0) $, and $ \hat{f}_j(t) =
\hat{F}_{2j-1}(t) $, $ j,k=1,2 $. The matrices $ {\bf U} $ and $
{\bf V} $ are derived as follows:
\begin{eqnarray} 
 & {\bf U} = \frac{1}{\mu} \left[ \begin{array}{cc}
  \beta c-\gamma s & -i\epsilon s \\
  -i\epsilon s & \beta c +\gamma s \end{array} \right],
  \hspace{1mm}
 {\bf V} = - \frac{i\kappa s}{\mu} \left[ \begin{array}{cc}
  0 & 1 \\ 1 & 0 \end{array} \right], &
\label{15}
\end{eqnarray}
where $ s \equiv \sin(\mu t) $ and $ c \equiv \cos(\mu t) $.

Similarly, we arrive at $ \langle \hat{\bf F}(t) \rangle = \langle
\hat{\bf F}^\dagger (t) \rangle = {\bf 0} $. On the other hand,
the correlation functions of the fluctuating operator forces $
\langle \hat{\bf F}(t)^{} \hat{\bf F}^{\dagger {\bf T}}(t)\rangle
$ at time $ t $ are nonzero:
\begin{eqnarray}  
 & \langle \hat{\bf F}(t)^{} \hat{\bf F}^{\dagger {\rm T}}(t)\rangle
  = \left[ \begin{array}{cc}
   {\bf F}_1(t) & {\bf F}_{12}(t) \\
   {\bf F}_{12}^{* {\bf T}}(t) & {\bf F}_2(t) \end{array} \right],  &
\label{16} \\
 & {\bf F}_1(t) = \frac{2\gamma}{\mu} \left( sc - \frac{\gamma
  s^2}{\mu} \right) \left[ \begin{array}{cc} 1 & 0 \\ 0 & 0 \end{array} \right]
  + \frac{\gamma\epsilon}{\mu^3} \left( sc -\mu t \right) {\bf F}_a , &
  \nonumber \\
 & {\bf F}_2(t) = \frac{2\gamma}{\mu} \left( sc + \frac{\gamma
  s^2}{\mu} \right) \left[ \begin{array}{cc} 0 & 0 \\ 0 & 1 \end{array} \right]
  + \frac{\gamma\epsilon}{\mu^3} \left( sc -\mu t \right) {\bf F}_a , &
  \nonumber \\
 & {\bf F}_{12}(t) = \frac{i\epsilon\gamma^2}{\mu^3} \left( sc- \mu t
  \right)  \left[ \begin{array}{cc} 1 & 0 \\ 0 & -1 \end{array} \right]
   + \frac{i\gamma}{\mu^2} s^2 \left[ \begin{array}{cc} \epsilon & -2\kappa \\ 0 & \epsilon \end{array}
  \right], &
  \nonumber \\
 & {\bf F}_a = \left[ \begin{array}{cc} -\epsilon &
  \kappa \\ \kappa & -\epsilon \end{array} \right]. & \nonumber
\end{eqnarray}

The comparison of formulas for the evolution matrices $ {\bf U}(t)
$ and $ {\bf V}(t) $ in Eq.~(\ref{15}) and the correlation
functions $ \langle \hat{\bf F}(t)^{} \hat{\bf F}^{\dagger {\bf
T}}(t)\rangle $ in Eq.~(\ref{16}) reveals a striking difference:
Whereas the evolution matrices behave periodically in time, the
correlation functions exhibit a linear-time dependence
superimposed on their otherwise periodic evolution. Detailed
investigations in Sec.~IV show that this property is responsible
for a gradual suppression of the nonclassicality in the system
evolution.

We note that there exist two platforms with $ \chi^{(2)} $
nonlinearity allowing for an experimental implementation of the
system with the Hamiltonian $ \hat{H} $ given in Eq.~(\ref{1}):
(i) nonlinear solid-state photonic structures \cite{PerinaJr2007a}
and (ii) nonlinearly interacting Rydberg atoms in cells
\cite{Boyer2008}. In both cases, photons are emitted or
annihilated in pairs in the process of parametric down-conversion
\cite{Mandel1995} or four-wave mixing \cite{Meystre2007} with
strong pumping. Considering the first platform, linear
corrugations at the surfaces of waveguiding structures allow for
the linear exchange of energy between two modes as well as they
can be used in principle to dissipate or actively amplify a mode
field. On the other side, additional atoms in their ground
(excited) states in resonance with the mode fields present in a
cell with nonlinearly interacting Rydberg atoms (e.g., rubidium,
\cite{Boyer2008}) cause damping (incoherent amplification) of the
mode fields using the second platform. As an experimental
realization of incoherent amplification is a difficult task,
passive $\mathcal{PT} $-symmetric non-Hermitian systems
\cite{Chimczak2023} may be considered to overcome this problem.

\section{Tailoring the reservoir properties}

The form of correlation functions $ \langle \hat{\bf F}(t)^{}
\hat{\bf F}^{\dagger {\bf T}}(t)\rangle $ depends on the
properties of the reservoir. There is the question whether a
suitable reservoir can be constructed such that the correlation
functions $ \langle \hat{\bf F}(t)^{} \hat{\bf F}^{\dagger {\bf
T}}(t)\rangle $ behave periodically and no loss of the
nonclassicality occurs for asymptotically long times.

As the relation between the fluctuating operator forces $ \hat{\bf
F}(t) $ and $ \hat{\bf L}(t) $ in Eq.~(\ref{6}) is linear, we may
invert the relation between their correlation functions in
Eq.~(\ref{9}). First, we rewrite Eq.~(\ref{9}) using Eq.~(\ref{8})
for the evolution matrix $ {\bf P} $:
\begin{eqnarray*} 
 \langle \hat{\bf F}(t)^{} \hat{\bf F}^{\dagger {\rm T}}(t')\rangle &=& \int_{0}^t
  d\tilde{t} \int_{0}^{t'}  d\tilde{t}' \;\;
  {\bf T}  \exp(\Lambda_{\bf M}\tilde{t}) {\bf T}^{-1} \nonumber \\
 & & \hspace{-20mm}  \times \langle \hat{\bf L}(t-\tilde{t})
  \hat{\bf L}^{\dagger {\rm T}}(t'-\tilde{t}')\rangle   {\bf T}^{-1 * T} \exp(\Lambda_{\bf M}\tilde{t}')
  {\bf T}^{* T}.
\end{eqnarray*}
Then, relying on the Markovian character of the fluctuating forces
$ {\bf L} $ in Eq.~(\ref{2}) and expressing the correlation
function matrix $ \langle \hat{\bf L}(t)\hat{\bf L}^{\dagger {\rm
T}}(t')\rangle $ as $ {\bf L}^0 \delta(t-t') $ we arrive at the
formula:
\begin{eqnarray} 
 \langle \hat{\bf F}(t)^{} \hat{\bf F}^{\dagger {\rm T}}(t)\rangle &=& \int_{0}^t
  d\tilde{t} {\bf T}  \exp(\Lambda_{\bf M}\tilde{t}) {\bf T}^{-1} {\bf L}^0  \nonumber \\
 & & \times {\bf T}^{-1 * T} \exp(\Lambda_{\bf M}\tilde{t})
  {\bf T}^{* T}.
\label{17}
\end{eqnarray}
Using inversion of Eq.~(\ref{17}) the following formula for the
correlation function matrix $ \langle \hat{\bf L}(t)^{} \hat{\bf
L}^{\dagger {\rm T}}(t')\rangle $ is obtained:
\begin{eqnarray} 
 \langle \hat{\bf L}(t)^{} \hat{\bf L}^{\dagger {\rm T}}(t')\rangle
   &=& {\bf L}^0 \delta(t-t'), \\
 {\bf L}^0 &=&
  {\bf T}  \exp(-\Lambda_{\bf M}t) {\bf T}^{-1} \frac{d}{dt} \langle \hat{\bf F}(t)
  \hat{\bf F}^{\dagger{\rm T}}(t)\rangle  \nonumber \\
 & & \times  {\bf T}^{-1 * T} \exp(-\Lambda_{\bf M}t)
  {\bf T}^{* {\rm T}}. \nonumber
\label{18}
\end{eqnarray}

Inserting the terms which are linearly proportional to time $ t $
in Eq.~(\ref{16}) into Eq.~(\ref{18}) we arrive at the
corresponding correlation function matrix:
\begin{eqnarray}   
 & \langle \hat{\bf L}^{t}(t) \hat{\bf L}^{t\dagger {\rm
   T}}(t')\rangle = \frac{\epsilon\gamma}{\mu^2} \left[ \begin{array}{cccc}
   \epsilon & -\kappa & -i\gamma & 0 \\
   -\kappa & \epsilon & 0 & i\gamma \\
   i\gamma & 0 & \epsilon & -\kappa \\
   0 & -i\gamma & -\kappa & \epsilon \end{array} \right]
   \delta(t-t') . & \nonumber \\
  & &
\label{19}
\end{eqnarray}

The reservoir correlation function matrix $ \langle \hat{\bf
L}^{\rm id}(t)\hat{\bf L}^{{\rm id} \dagger {\rm T}}(t')\rangle $
that guarantees a periodic evolution of the system, and, thus,
does not lead to nonclassicality deterioration, can then be
written with the help of Eq.~(\ref{19}) as:
\begin{eqnarray}   
 &\langle \hat{\bf L}^{\rm id}(t)\hat{\bf L}^{{\rm id} \dagger {\rm
   T}}(t')\rangle = \frac{\epsilon\gamma}{\mu^2} \left[ \begin{array}{cccc}
   2\mu^2 /\epsilon -\epsilon & \kappa & i\gamma & 0 \\
   \kappa & -\epsilon & 0 &  -i\gamma \\
   -i\gamma & 0 & -\epsilon & \kappa \\
   0 & i\gamma & \kappa & 2\mu^2 /\epsilon -\epsilon \end{array} \right]
   & \nonumber \\
 & \times \delta(t-t') . &
\label{20}
\end{eqnarray}

\begin{figure}  
 \includegraphics[width=0.9\hsize]{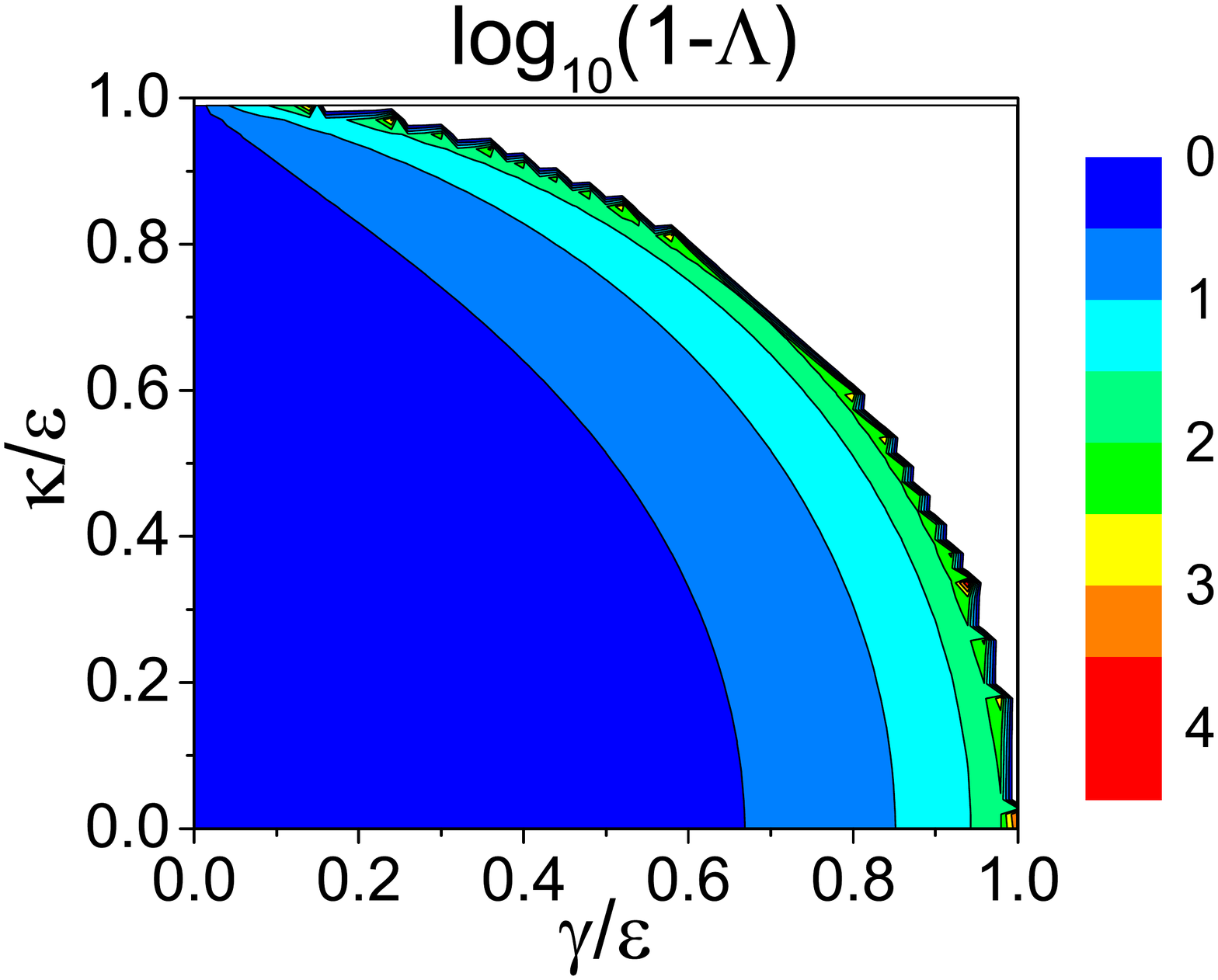}
 \caption{Strength $ \Lambda $ of the sink versus the
  damping rate $ \gamma $ and the coupling strength $ \kappa $ in units of the
 exchange energy rate~$ \epsilon $.}
\label{fig1}
\end{figure}

The matrix in Eq.~(\ref{20}) has two doubly degenerated
eigenvalues $ \nu_{\pm} $:
\begin{equation}   
 \nu_{\pm} = \frac{\gamma}{\mu^2} \left( -\kappa^2 - \gamma^2 \pm
  \sqrt{ \mu^4 + \epsilon^2(\kappa^2 + \gamma^2) } \right).
\label{21}
\end{equation}
The eigenvalue $ \nu_{-} $ is negative for $ \mu > 0 $, i.e., in
the region with the periodic behavior of the system. We have $ \mu
= 0 $ at EPs and so $ \nu_+ \rightarrow \gamma $ and $ \nu_-
\rightarrow -\infty $. This means that the reservoir with the
correlation function matrix $ \langle \hat{\bf L}^{\rm
id}(t)\hat{\bf L}^{{\rm id} \dagger {\rm T}}(t')\rangle $ has the
property of sink models that take energy from the system. The
strength of such a sink can be quantified using parameter $
\Lambda $ defined as the sum of all real eigenvalues in the area $
\mu \ge 0 $:
\begin{equation}   
 \Lambda = 2( \nu_+ + \mu_- ) .
\label{22}
\end{equation}
Substituting Eq.~(\ref{21}) into Eq.~(\ref{22}), we arrive at the
formula
\begin{equation}   
 \Lambda = 4\gamma \left( 1 - \frac{\epsilon^2}{\mu^2} \right).
\label{23}
\end{equation}
According to Eq.~(\ref{23}), the closer to an EP the system
parameters are, the more negative the sink strength $ \Lambda $
is, as shown in Fig.~\ref{fig1}. It even goes to $ -\infty $ at an
EP.

The correlation function matrix $ \langle \hat{\bf L}^{\rm
id}(t)\hat{\bf L} ^{{\rm id} \dagger {\rm T}}(t')\rangle $ in
Eq.~(\ref{20}) describes two coupled oscillators. Neglecting their
coupling, both oscillators have identical eigenfrequencies
expressed as $ \tilde \nu_{\pm} = \gamma ( -\kappa^2 - \gamma^2
\pm \sqrt{ \mu^4 + \epsilon^2\kappa^2} ) / \mu^2 $. At least one
eigenfrequency is negative and their sum gives the sink parameter
$ \Lambda $ written in Eq.~(\ref{23}). We note that, provided that
the diagonal elements of the correlation function matrices of
these oscillators were nonnegative, they describe the squeezed
reservoirs \cite{Gardiner1985}. If the squeezed reservoirs are
considered the system dynamics qualitatively changes from the
point of view of nonclassical and entangled states generation
\cite{Dum1992,Munro1995}. Such states are obtained even for long
times owing to the reservoir nonclassicality that is constantly
being transferred into the system
\cite{Dum1992,Munro1995,Kowalewska2010,Kowalewska2019,Manzano2018}.
We also note that the nonclassical and entangled states emerge
when nonlinear interactions with reservoirs are taken into account
\cite{Gilles1994,Everitt2014}.

A usual physical reservoir is composed of the populated modes
whose random influence to the system compensates for the system
loss (damping) or gain (amplification) of energy during its
evolution. This means that there are nonnegative eigenvalues $
\nu_{\pm} $, similarly as in the case of the original correlation
functions in Eq.~(\ref{2}) [$ \nu_+ = 2\gamma $, $ \nu_- = 0 $].
Thus, the terms in the correlation function matrix $ \langle
\hat{\bf F}(t)^{} \hat{\bf F}^{\dagger {\rm T}}(t)\rangle $ in
Eq.~(\ref{16}), which are linearly proportional to time $ t $,
cannot be compensated by a suitable physical reservoir and, thus,
the nonclassicality deterioration occurs inevitably during the
evolution of the quantum $ \mathcal{PT} $-symmetric system. This
reflects the fundamental fact that whereas damping and
amplification can compensate each other in (semi)classical
coherent dynamics, the effects of fluctuating forces accompanying
damping and amplification cannot be mutually suppressed.

Moreover, the complete omission of the Langevin forces $ \hat{l}_j
$ and $ \hat{l}_j^\dagger $, $ j=1,2 $, results in a nonphysical
behavior that violates the fluctuation-dissipation theorem. This
situation corresponds to the semiclassical model described by the
non-Hermitian Hamiltonian
\begin{eqnarray}  
 \hat{H}^{\rm sc} &=& -i\gamma \hat{a}_1^\dagger\hat{a}_1 + i\gamma
  \hat{a}_2^\dagger\hat{a}_2 + \left[ \epsilon \hat{a}_1^\dagger\hat{a}_2 +
   \kappa \hat{a}_1\hat{a}_2 + {\rm h.c.} \right]. \nonumber \\
  & &
\label{24}
\end{eqnarray}
We show in Secs.~IV and V that the declinations of the system
evolution from the physical one are qualitatively similar for both
models.

\section{Nonclassicality and entanglement in $ \mathcal{PT}
$-symmetric systems with different levels of reservoir
fluctuations}

A detailed role of the reservoir fluctuating forces in the
evolution of nonclassical properties of two coupled oscillator
modes is elucidated considering three models differently including
reservoir fluctuations: (1) a physically-consistent model fully
including reservoir fluctuations; (2) an ideal (sink) model with a
partial inclusion of reservoir fluctuations obeying the
fluctuation-dissipation relations and giving a periodic solution,
and (3) a semiclassical model with no reservoir fluctuations thus
violating the fluctuation-dissipation theorem.

We directly compare these three models for arbitrary values of all
coefficients of the normal characteristic function in
Eq.~(\ref{25}) below as well as by considering various properties
of the modes. For simplicity, we restrict our attention to the
initial coherent states in both oscillator modes. These states are
Gaussian and they remain Gaussian during the evolution owing to
the linear Heisenberg-Langevin equations in Eq.~ (\ref{3}). Their
normal characteristic function $ C_{\mathcal N} $ can be expressed
as follows~\cite{Perina1991}:
\begin{eqnarray} 
 C_{\cal N}(\mu_1,\mu_2,t) &=& \exp\Biggl\{ \sum_{j=1,2} \Bigl[
   \left( \alpha_j^*(t)\mu_j - {\rm c.c.}\right)
   \nonumber\\
 & & \hspace{-5mm} \mbox{} -B_j(t)|\mu_j|^2 + \left(C_j(t)\mu_j^{2*} + {\rm c.c.}\right)/2 \Bigr]
  \nonumber \\
 & & \hspace{-5mm} \mbox{} + \left[ D(t)\mu_1^*\mu_2^* + \bar{D}(t)\mu_1\mu_2^* + {\rm
   c.c.}\right] \Bigr\},
\label{25}
\end{eqnarray}
where c.c. stands for the complex conjugated term. Definitions of
the time dependent parameters occurring in Eq.~(\ref{25}), as well
as their simplified forms valid for the initial coherent states,
are given as follows:
\begin{eqnarray}  
 B_j(t) &\equiv& \langle\delta\hat{a}_j^\dagger(t)\delta\hat{a}_j(t)\rangle
  = \sum_{l=1,2} \left[ |V_{jl}(t)|^2 +
  \langle\hat{f}_j^\dagger(t)\hat{f}_j(t)\rangle \right], \nonumber \\
 C_j(t) &\equiv& \langle [\delta\hat{a}_j(t)]^2\rangle
  = \sum_{l=1,2} \left[ U_{jl}(t) V_{jl}(t) +
  \langle[\hat{f}_j(t)]^2\rangle \right], \nonumber \\
 D(t) &\equiv& \langle\delta\hat{a}_1(t)\delta\hat{a}_2(t)\rangle
   \nonumber \\
 & & = \sum_{l=1,2} \left[ U_{1l}(t)V_{2l}(t) +
  \langle\hat{f}_1(t)\hat{f}_2(t)\rangle \right], \nonumber \\
 \bar{D}(t) &\equiv& -\langle\delta\hat{a}_1^\dagger(t)\delta\hat{a}_2(t)\rangle
  \nonumber \\
 & &  = -\sum_{l=1,2} \left[V_{1l}^*(t)V_{2l}(t)  +
   \langle\hat{f}_1^\dagger(t)\hat{f}_2(t)\rangle \right],
\label{26}
\end{eqnarray}
where $ \delta \hat{a}_j = \hat{a}_j - \langle \hat{a}_j \rangle $
for $ j=1,2 $.

We quantify the nonclassicality of the system using the Lee
nonclassicality depth $ \tau $ \cite{Lee1991} derived from the
threshold value $ s_{\rm th} $ of the field-operator ordering
parameter at which the corresponding quasi-distribution $ \Phi_s $
of field amplitudes starts to behave as a classical function:
\begin{equation}  
 \tau = \frac{1-s_{\rm th}}{2} .
\label{27}
\end{equation}
To arrive at the nonclassicality depth $ \tau $, we first
determine the characteristic function $ C_{s} $ for an arbitrary
ordering parameter $ s $. The function $ C_{s} $ keeps the
Gaussian form of the normal characteristic function $ C_{\cal N} $
with the following modified parameters \cite{Perina1991}:
\begin{equation} 
 C_{s}(\mu_1,\mu_2,t) = \left. C_{\cal N}(\mu_1,\mu_2,t) \right|_{
 B_j \leftarrow B_{j,s} = (1-s)/2 + B_j, j=1,2}.
\label{28}
\end{equation}
The quasi-distribution $ \Phi_s $ associated to the characteristic
function $ C_{s} $ in Eq.~(\ref{28}) is obtained by the following
Fourier transform:
\begin{eqnarray}  
 \Phi_{s}(\alpha_1,\alpha_2,t) &=& \frac{1}{\pi^2} \prod_{j=1}^2
   \int d^2\mu_j \exp(\alpha_j\mu_j^* - \alpha_j^*\mu_j)\nonumber \\
  & &   \times C_{s}(\mu_1,\mu_2,t).
\label{29}
\end{eqnarray}
The existence of quasi-distribution $ \Phi_s $ as an ordinary
function requires a nonnegative determinant of the matrix $ {\bf
K}_{\Phi_s} $ of coefficients of the complex quadratic form
occurring in the argument of the exponential function on the
r.h.s. of Eq.~(\ref{28}):
\begin{equation}   
 {\bf K}_{\Phi_s} = \frac{1}{2} \left[ \begin{array}{cccc}
   -B_{1,s} & C_1^* & \bar{D}^* & D \\
   C_1 &  -B_{1,s} & D^* & \bar{D} \\
   \bar{D} & D & -B_{2,s} & C_2^* \\
   D^* & \bar{D}^* & C_2 &  -B_{2,s} \end{array} \right].
\label{30}
\end{equation}
For classical distributions $ \Phi_s $ occurring for $ s\le s_{\rm
th} $, all the four eigenvalues of the matrix $ {\bf K}_{\Phi_s} $
are negative, which results in its positive determinant. At $
s=s_{\rm th} $, one of these eigenvalues is zero and becomes
positive for $ s> s_{\rm th} $. Taking into account that the
diagonal elements of the matrix $ {\bf K}_{\Phi_{s_{\rm th}}} $
are given as $ B_{j,s_{\rm th}} = (1-s_{\rm th})/2 + B_j = \tau +
B_j $, $ j=1,2 $, the nonclassicality depth $ \tau $ is given as
the greatest positive eigenvalue of the matrix $ {\bf
K}_{\Phi_{\cal N}} \equiv {\bf K}_{\Phi_{s=1}} $. Applications of
these results can be found, e.g., in
\cite{Arkhipov2015,Arkhipov2016,Arkhipov2016a}.

Applying this procedure to the characteristic function $
C_{j,s}(\mu_j,t) $ of mode $ j $, we easily derive the following
formula for the corresponding nonclassicality depth $ \tau_j $:
\begin{equation}   
 \tau_j = {\rm max}\{ 0,|C_j| - B_j\}.
\label{31}
\end{equation}

Entanglement represents arguably the most striking manifestation
of nonclassicality. Logarithmic negativity $ E_N $
\cite{Horodecki2009} is usually used to quantify it. For a
two-mode Gaussian field with the characteristic function $ C_{\cal
N} $ given in Eq.~(\ref{25}), the negativity $ E_N $ is determined
from the coherence matrix $ \sigma^{PT} $ belonging to the system
with partially transposed mode 2 and, thus, defined for the vector
$ (\hat{q}_1,\hat{p}_1,\hat{q}_2,-\hat{p}_2) $ \cite{Adesso2007}:
\begin{eqnarray}  
 {\bf \sigma}^{PT} &=& \left[ \begin{array}{cc}
  {\bf \sigma}_1 &  {\bf \sigma}_{12}^{PT} \\
  \left[{\bf \sigma}_{12}^{PT}\right]^{T} &  {\bf \sigma}_2^{PT} \end{array}
  \right] ,
\label{32} \\
  {\bf \sigma}_1 &=& \left[ \begin{array}{cc}
   1+2B_1+2\Re\{C_1\} & 2\Im\{C_1\} \\
   2\Im\{C_1\} &   1+2B_1-2\Re\{C_1\} \end{array}
  \right] , \nonumber \\
  {\bf \sigma}_2^{PT} &=& \left[ \begin{array}{cc}
   1+2B_2+2\Re\{C_2\} & -2\Im\{C_2\} \\
   -2\Im\{C_2\} &   1+2B_2-2\Re\{C_2\} \end{array}
  \right] , \nonumber \\
  {\bf \sigma}_{12}^{PT} &=& 2\left[ \begin{array}{cc}
   \Re\{D-\bar{D}\} & \Im\{-D+\bar{D}\} \\
   \Im\{D+\bar{D}\} & \Re\{D+\bar{D}\} \end{array}
  \right] \nonumber ,
\end{eqnarray}
where symbol $ T $ stands for the transposed matrix. The partially
transposed coherence matrix  $ \sigma^{PT} $ has two symplectic
eigenvalues $ \nu_{\pm} $ defined in terms of the invariants $
\Delta = {\rm det}\{ {\bf \sigma}^{PT} \} $ and $ \delta = {\rm
det}\{ {\bf \sigma}_1\} + {\rm det}\{ {\bf \sigma}_2^{PT}\} +
2{\rm det}\{ {\bf \sigma}_{12}^{PT}\} $ \cite{Adesso2007}:
\begin{equation}  
 \nu_{\pm} = \sqrt{ \frac{\delta}{2} \pm \sqrt{ \frac{\delta^2}{4}
 - \Delta }}.
 \label{33}
\end{equation}
The symplectic eigenvalue $ \nu_- $ then determines the negativity
$ E_N $ along the formula:
\begin{equation}  
 E_N = {\rm max}\{0,-\ln(\nu_-)\}.
\label{34}
\end{equation}

Considering the initial vacuum state in both modes and
substituting the solution to the Heisenberg-Langevin equations
given in Eqs.~(\ref{14})---(\ref{16}) into Eqs.~(\ref{26}) for the
coefficients of the normal characteristic function $ C_{\cal N} $,
we arrive at the following formulas:
\begin{eqnarray} 
 B_1(t) &=& \frac{\kappa^2}{2\mu^2} - \frac{\kappa^2}{2\mu^2} {\rm c}(t) -
 \frac{\epsilon^2\gamma}{2\mu^3} {\rm s}(t) +
 \frac{\epsilon^2\gamma}{\mu^2}t , \nonumber \\
 B_2(t) &=& \frac{\kappa^2+2\gamma^2}{2\mu^2} - \frac{\kappa^2+2\gamma^2}{2\mu^2}{\rm c}(t) +
 \frac{\epsilon^2\gamma}{2\mu^3} {\rm s}(t) +
 \frac{\epsilon^2\gamma}{\mu^2}t , \nonumber \\
 C_1(t) &=& C_2(t) = - \frac{\epsilon\kappa}{2\mu^2} + \frac{\epsilon\kappa}{2\mu^2}{\rm c}(t) +
 \frac{\epsilon\kappa\gamma}{2\mu^3} {\rm s}(t) -
 \frac{\epsilon\kappa\gamma}{\mu^2}t , \nonumber \\
 iD(t) &=& \frac{\kappa\gamma}{2\mu^2} - \frac{\kappa\gamma}{2\mu^2}{\rm c}(t)
  + i\frac{\kappa}{2\mu} {\rm s}(t) , \nonumber \\
  i\bar{D}(t) &=& \frac{\epsilon\gamma}{2\mu^2} - \frac{\epsilon\gamma}{2\mu^2}{\rm c}(t)
  - \frac{\epsilon\gamma^2}{2\mu^3} {\rm s}(t) + \frac{\epsilon\gamma^2}{\mu^2}t ,
\label{35}
\end{eqnarray}
where $ {\rm s}(t) \equiv \sin(2\mu t) $ and $ {\rm c}(t) \equiv
\cos(2\mu t) $. The terms linearly proportional to time $ t $ are
apparent in Eq.~(\ref{35}). They disappear when the ideal
reservoir is assumed. We note that similar linear time dependence
of some physical quantities was observed in
\cite{Purkayastha2020}.

For comparison, we write the above coefficients for the
semiclassical model in which the fluctuating forces are completely
neglected:
\begin{eqnarray} 
 B_1^{\rm sc}(t) &=& B_2^{\rm sc}(t) = \frac{\kappa^2}{2\mu^2}[1-{\rm c}(t)] , \nonumber \\
 C_1^{\rm sc}(t) &=& C_2^{\rm sc}(t) = - \frac{\epsilon\kappa}{2\mu^2} [1-{\rm c}(t)], \nonumber \\
 D^{\rm sc}(t) &=& i \frac{\kappa\gamma}{2\mu^2} [1-{\rm c}(t)], \nonumber \\
  \bar{D}^{\rm sc}(t) &=& 0 .
\label{36}
\end{eqnarray}

To assess the evolution of nonclassicality and entanglement, we
determine the maximal values of the nonclassicality depths $ \tau
$ and the negativity $ E_N $ in the first period of the periodical
solutions found in the ideal (sink) model including partial
reservoir fluctuations and the semiclassical model with no
reservoir fluctuations:
\begin{eqnarray}   
 &\tau^{\rm max} = \max_{t\in\langle 0,2\pi/|\mu|\rangle} \{\tau(t)\} ,
  & \nonumber \\
 & E_N^{\rm max} = \max_{t\in\langle 0,2\pi/|\mu|\rangle} \{ E_N(t) \}. &
\label{37}
\end{eqnarray}

The extremal quantities defined in Eq.~(\ref{37}) are reasonable
also for the physically-consistent model including complete
reservoir fluctuations as, during the evolution, the level of
noise in the system increases, which gradually conceals both
nonclassicality and entanglement. Moreover, in the long-time limit
of this model, the terms linearly proportional to time $ t $
prevail. This results in a considerable simplification of the
coefficients in Eq.~(\ref{36}):
\begin{eqnarray} 
 B_1^\infty (t) &=& B_2^\infty (t) = \frac{\epsilon^2\gamma}{\mu^2}t , \nonumber \\
 C_1^\infty(t) &=& C_2^\infty(t) = -\frac{\epsilon\kappa\gamma}{\mu^2}t , \nonumber \\
 D^\infty(t) &=& 0  , \nonumber \\
 \bar{D}^\infty(t) &=& -i\frac{\epsilon\gamma^2}{\mu^2}t .
\label{38}
\end{eqnarray}
Substitution of the long-time formulas in~(\ref{38}) into the
matrix $ {\bf K}_{\Phi_{\cal N}} $ leads to nonpositive
eigenvalues. We note that the greatest eigenvalue is doubly
degenerated and it is given by the formula $ - B_1^\infty + \sqrt{
|C_1^\infty|^2 + |\bar{D}^\infty|^2} $. This means that the states
are classical. Equations (\ref{31}) and (\ref{34}) for the
nonclasicality depths $ \tau_j $ of the modes and negativity $ E_N
$, respectively, confirm this:
\begin{eqnarray}   
 &\tau_1^\infty(t) = \tau_2^\infty(t) = \tau^\infty(t) = 0,
  & \nonumber \\
 & E_N^\infty(t) = 0. &
\label{39}
\end{eqnarray}

\begin{figure*}[t]  
 \includegraphics[width=0.225\hsize]{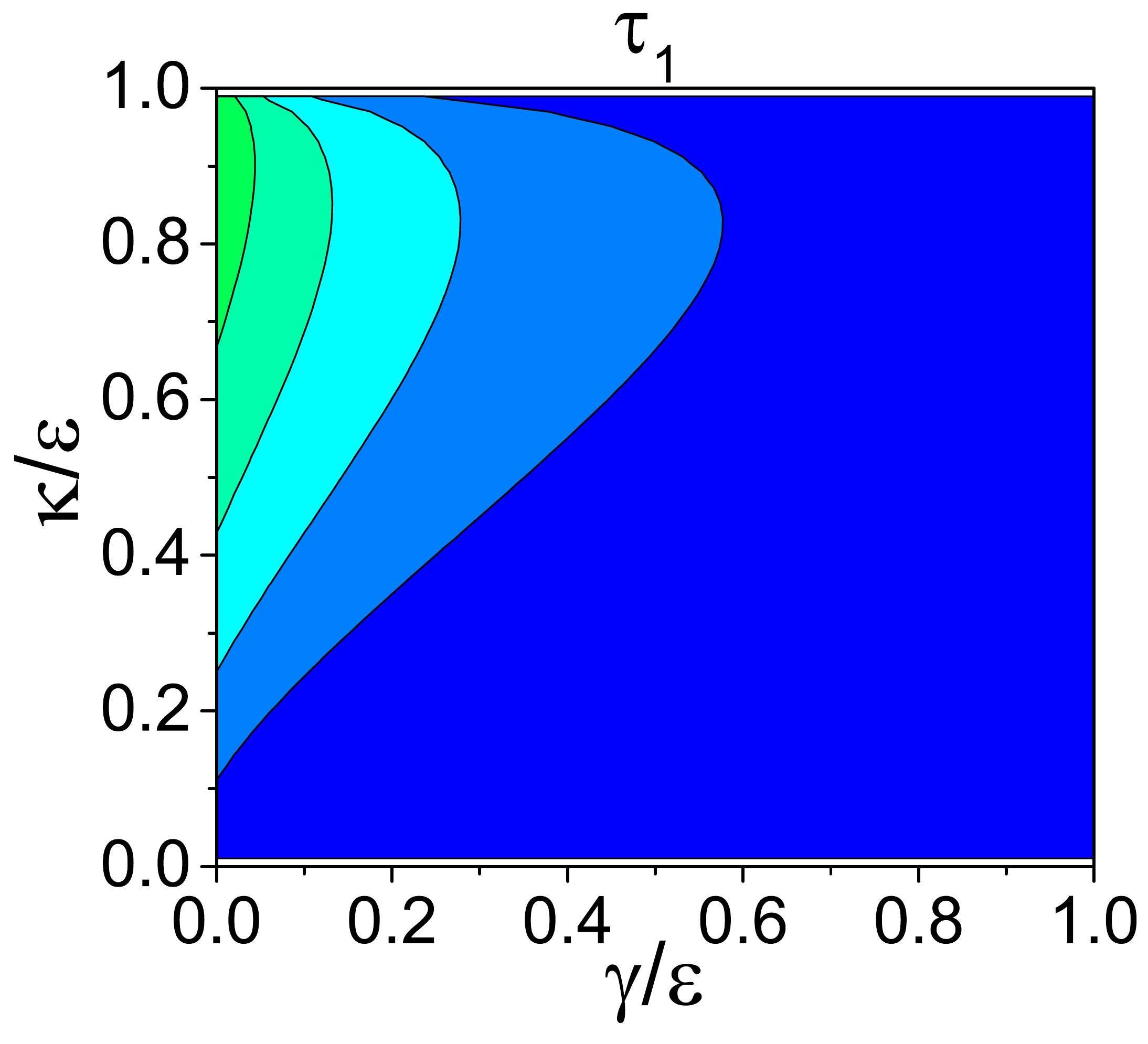}
 \includegraphics[width=0.225\hsize]{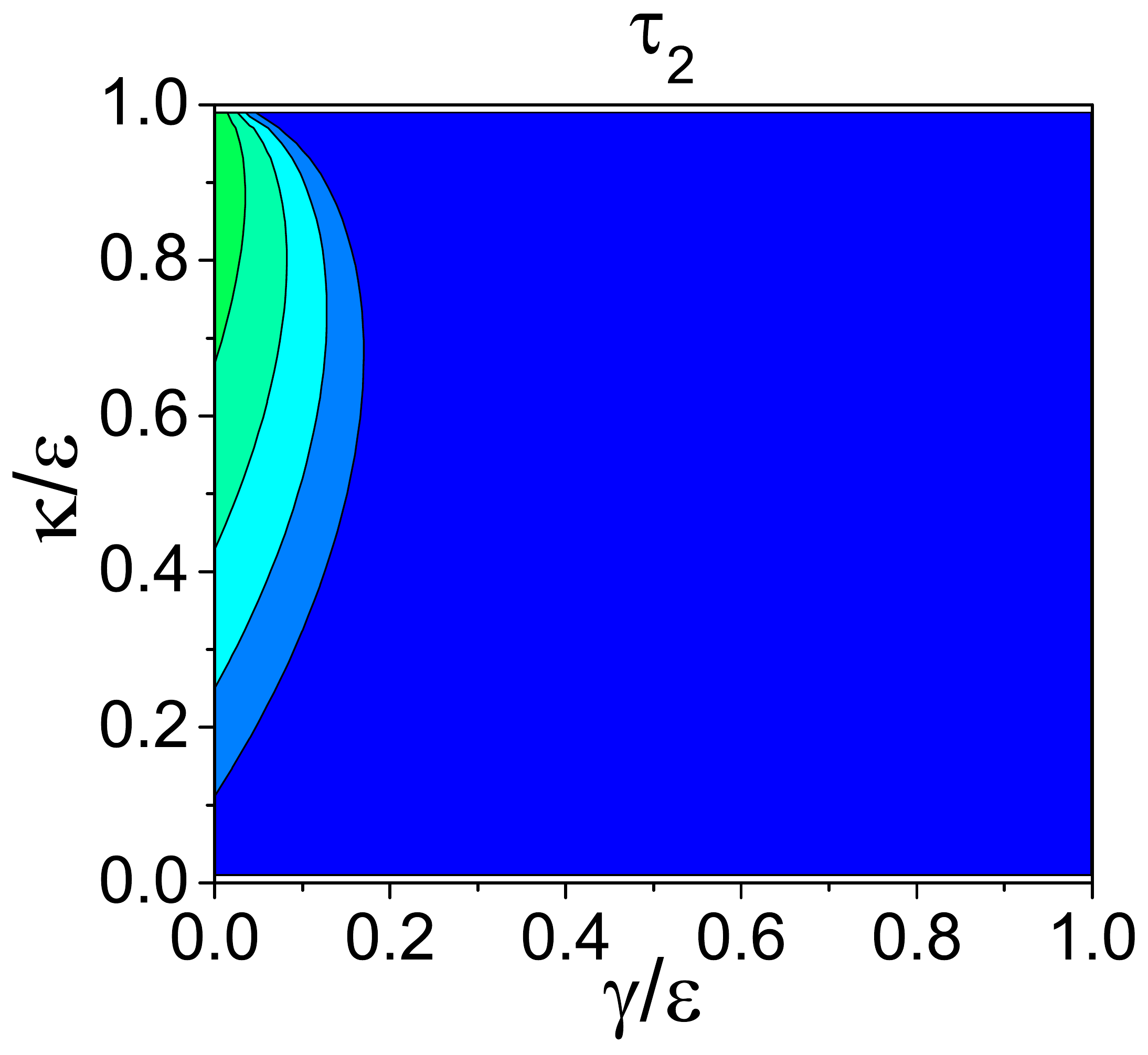}
 \includegraphics[width=0.26\hsize]{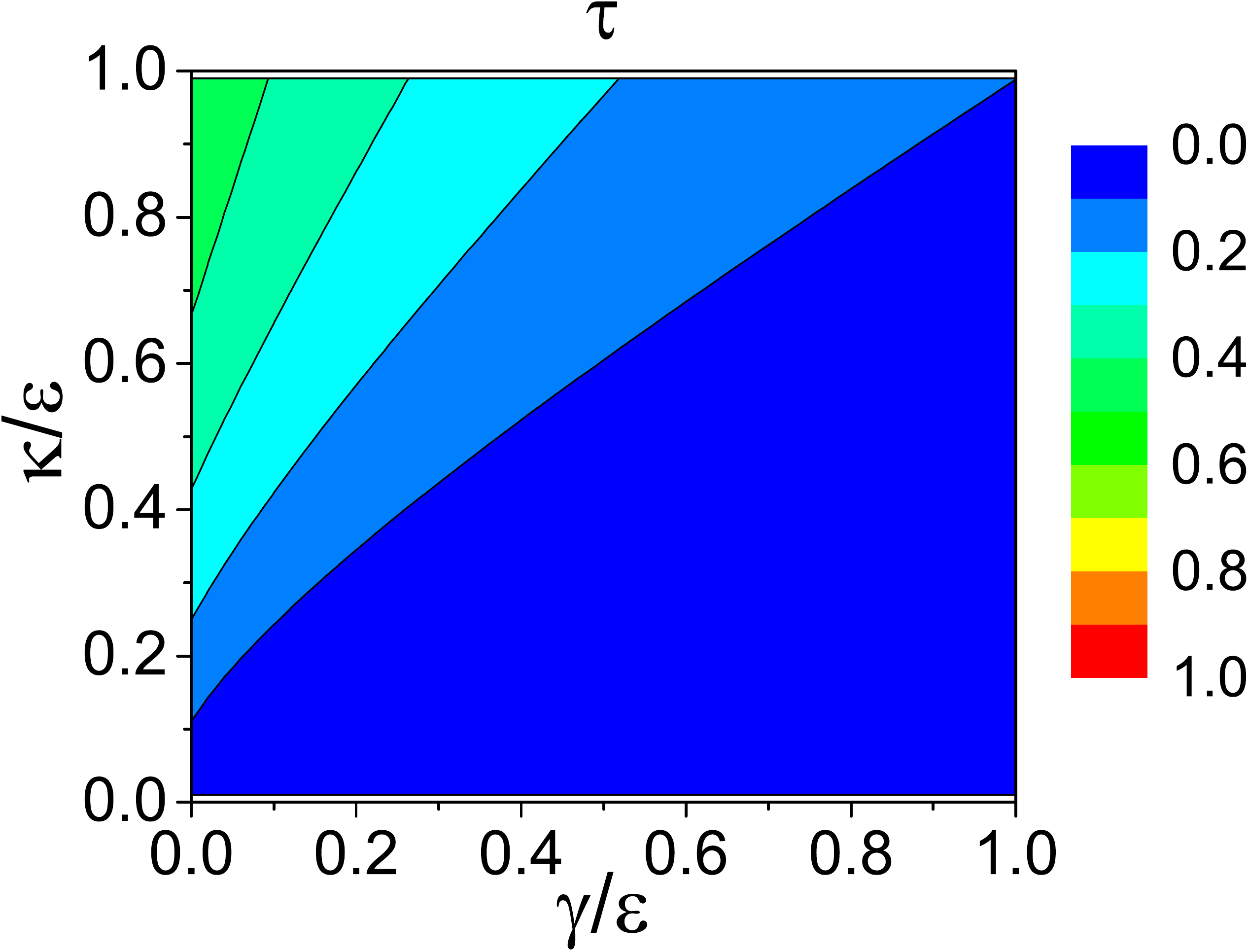}
 \includegraphics[width=0.26\hsize]{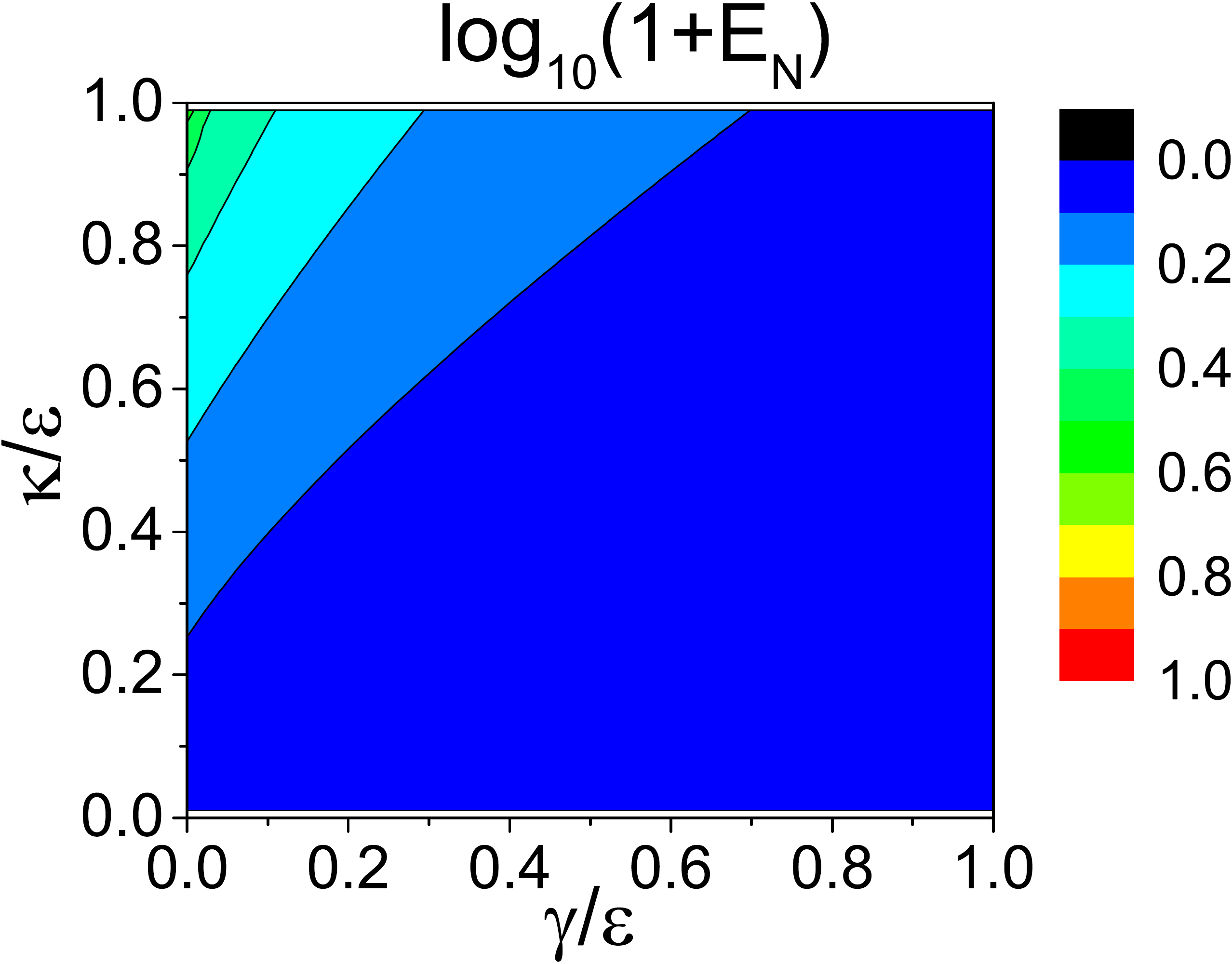}
 \vspace{2mm}
 \centerline{ \small (a) \hspace{.22\hsize} (b) \hspace{0.22\hsize} (c) \hspace{.22\hsize} (d)}

 \includegraphics[width=0.225\hsize]{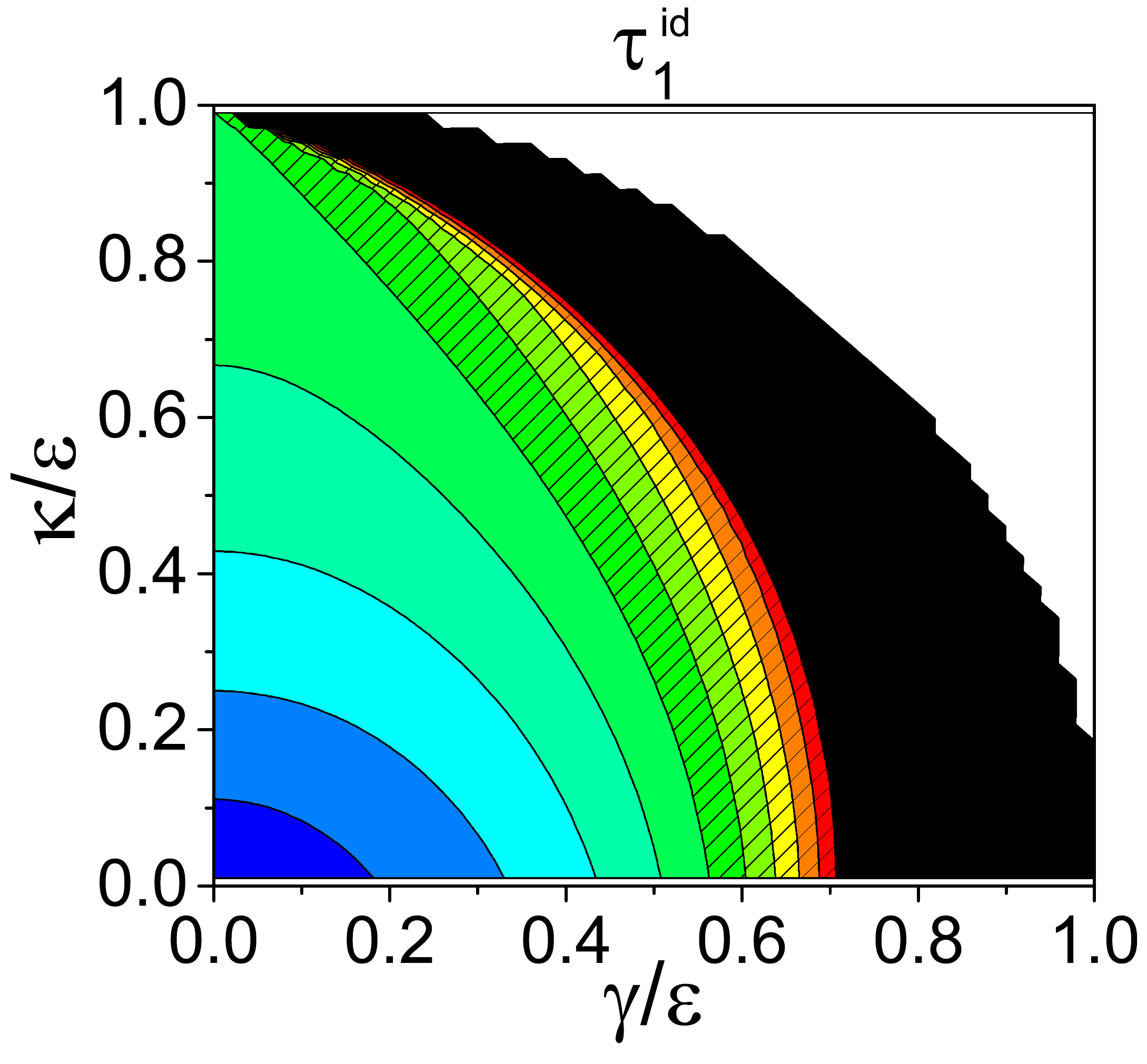}
 \includegraphics[width=0.225\hsize]{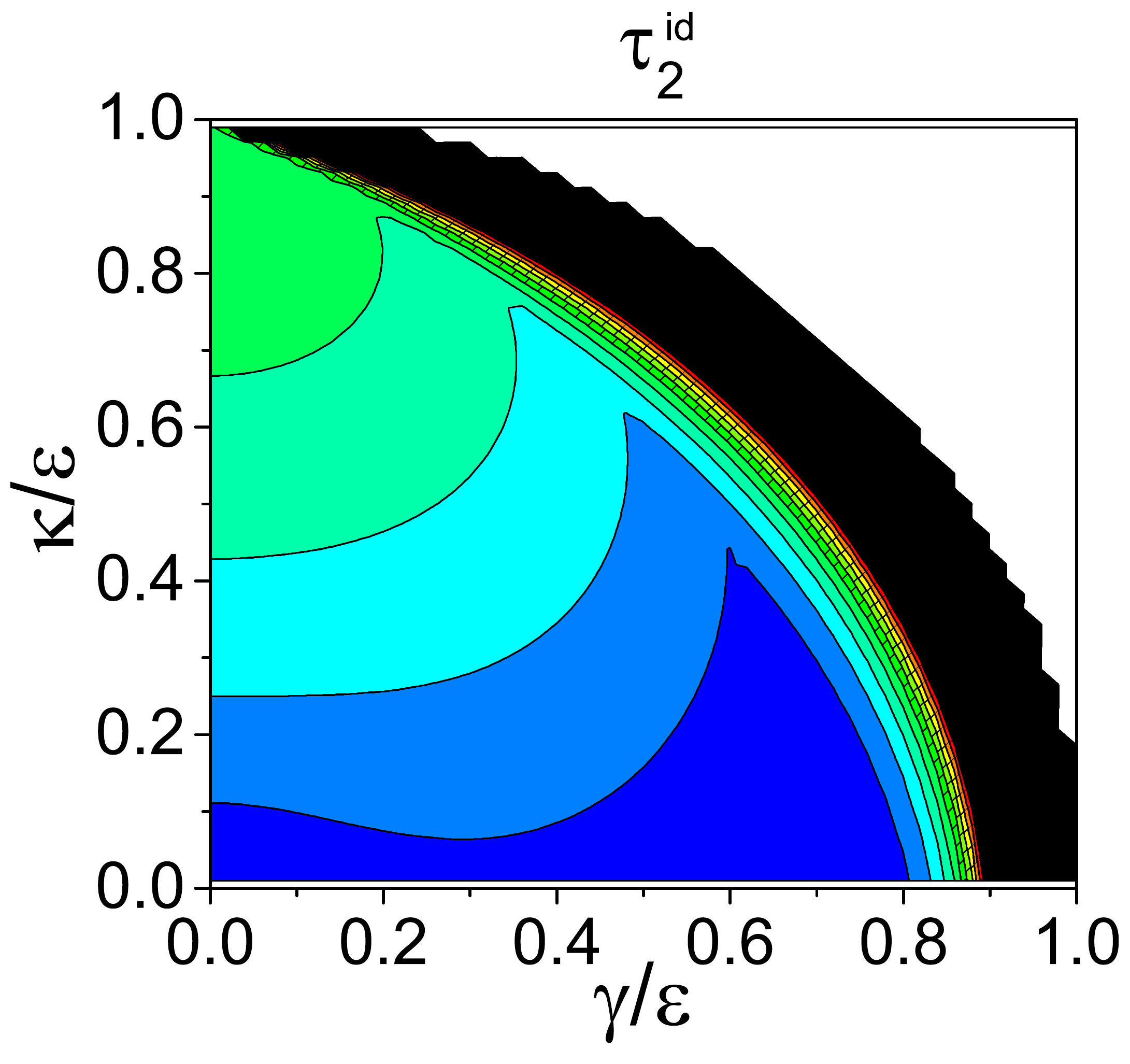}
 \includegraphics[width=0.26\hsize]{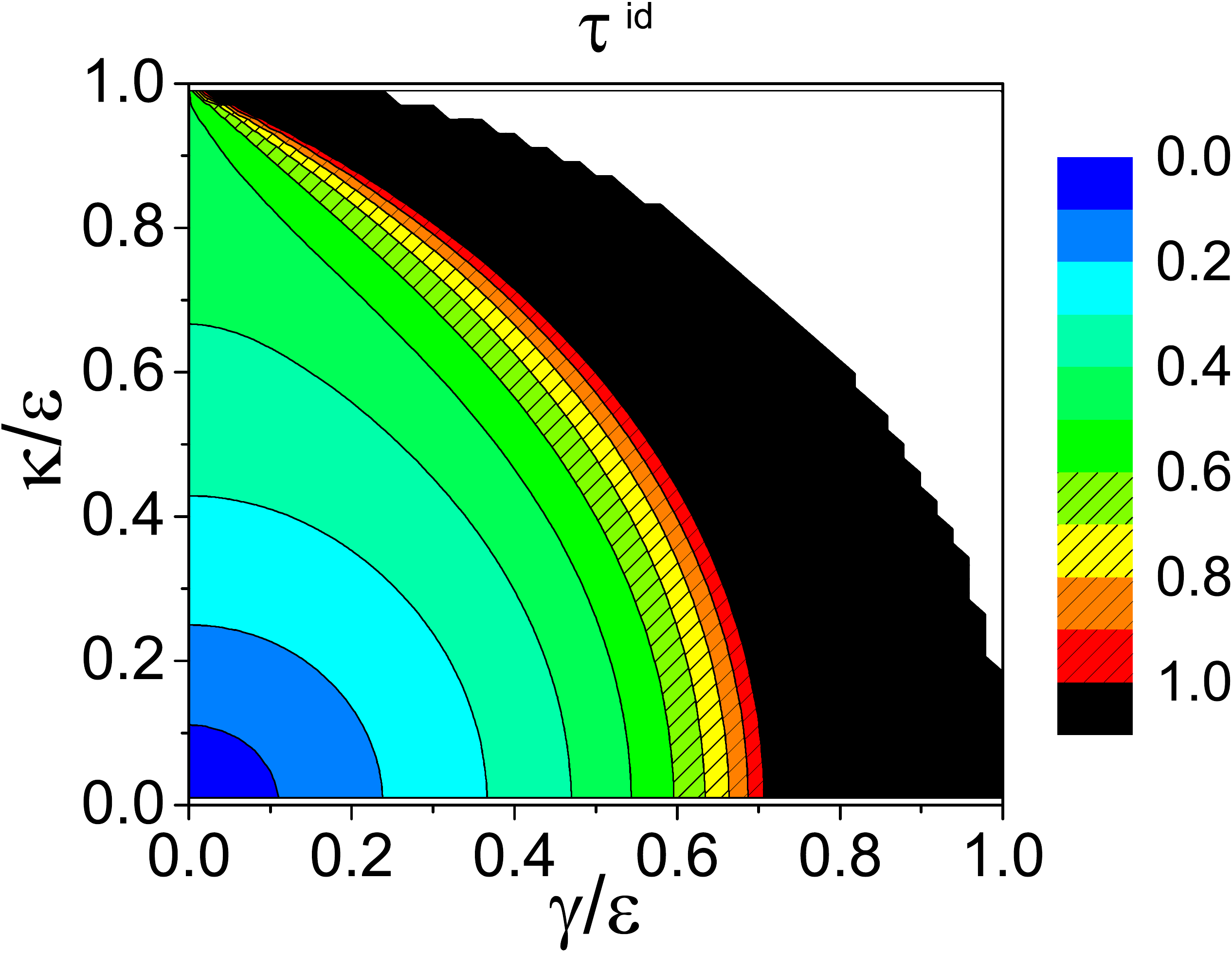}
 \includegraphics[width=0.26\hsize]{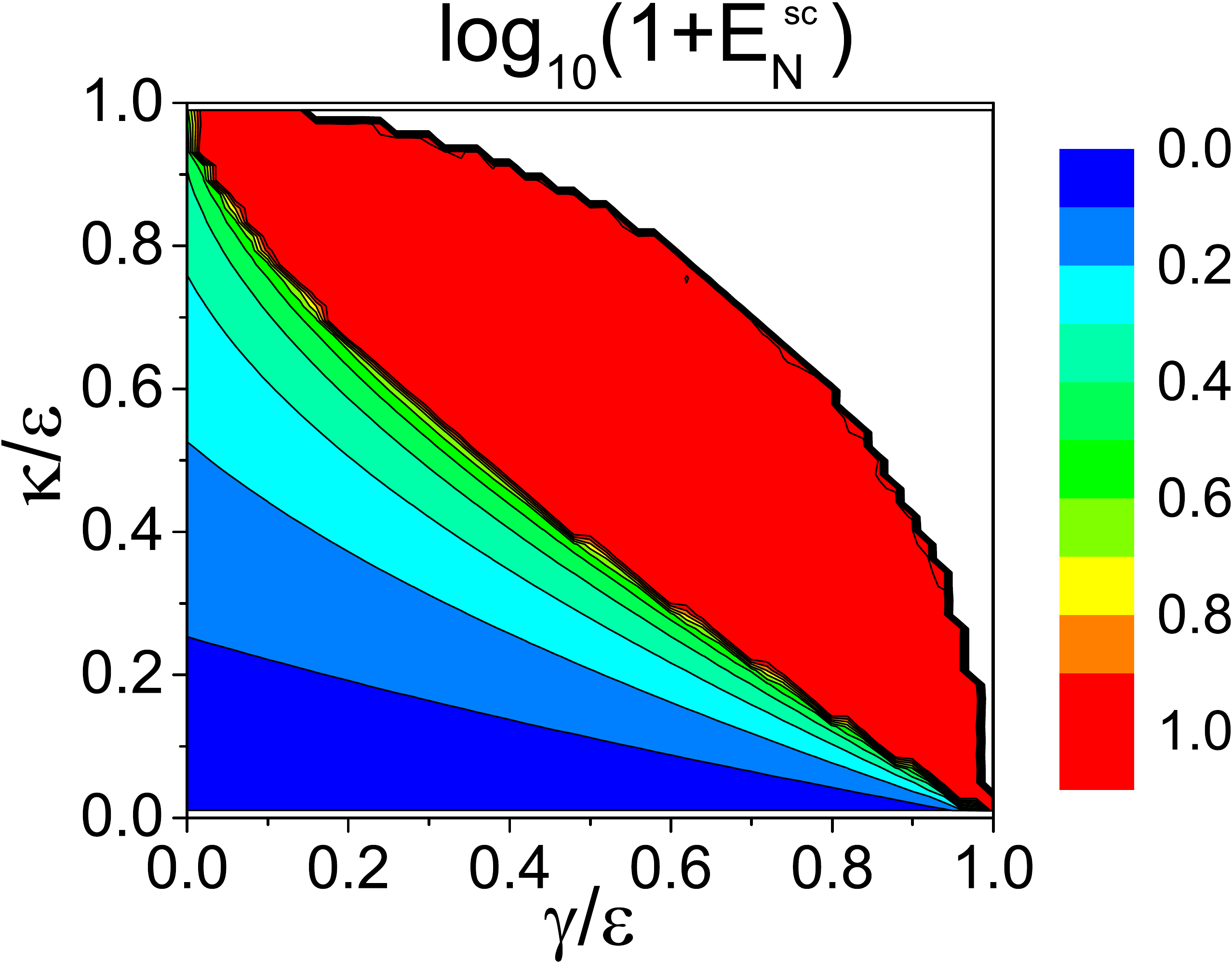}
 \vspace{2mm}
 \centerline{ \small (e) \hspace{.22\hsize} (f) \hspace{0.22\hsize} (g) \hspace{.22\hsize} (h)}

 \includegraphics[width=0.225\hsize]{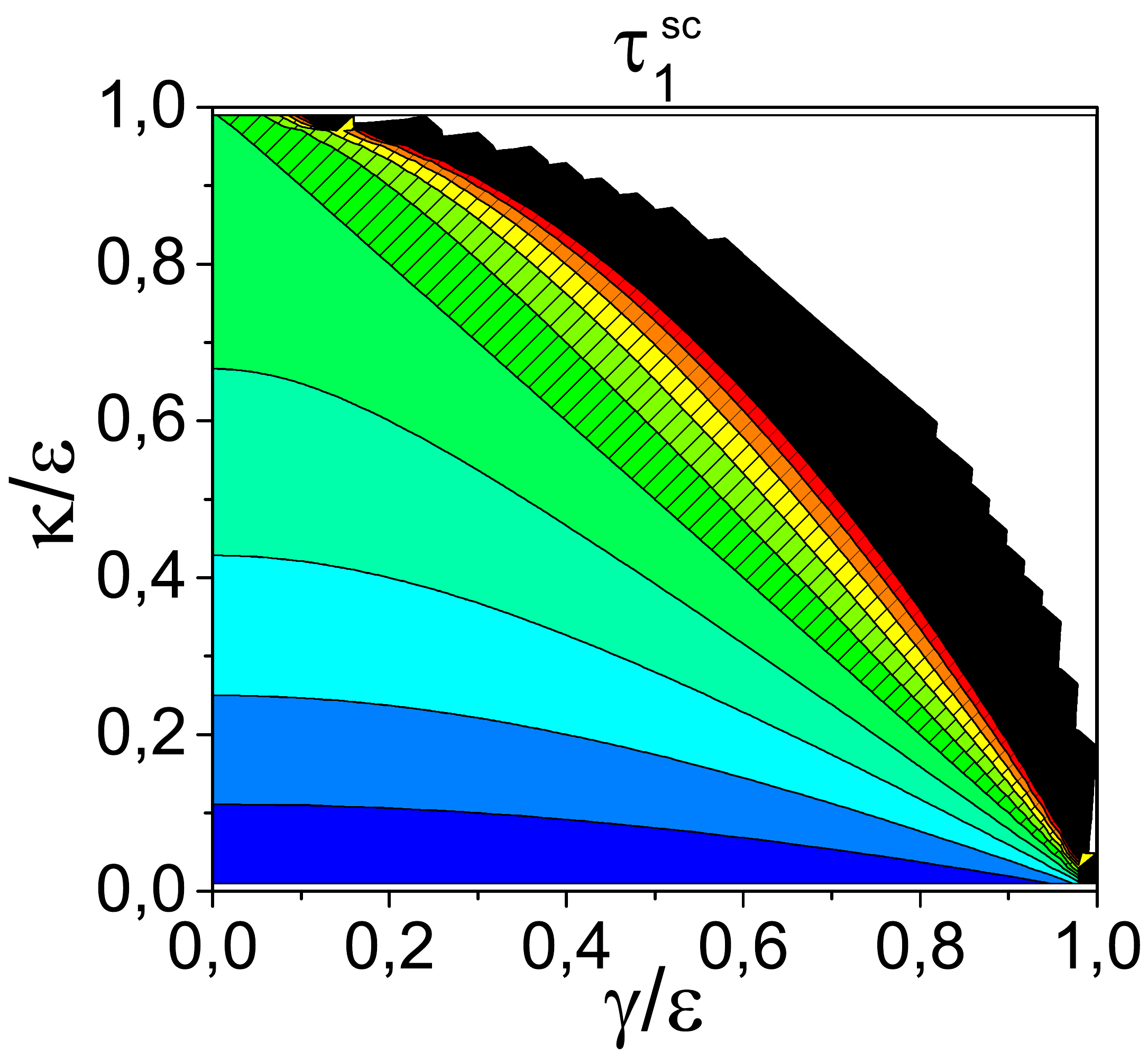}
 \includegraphics[width=0.225\hsize]{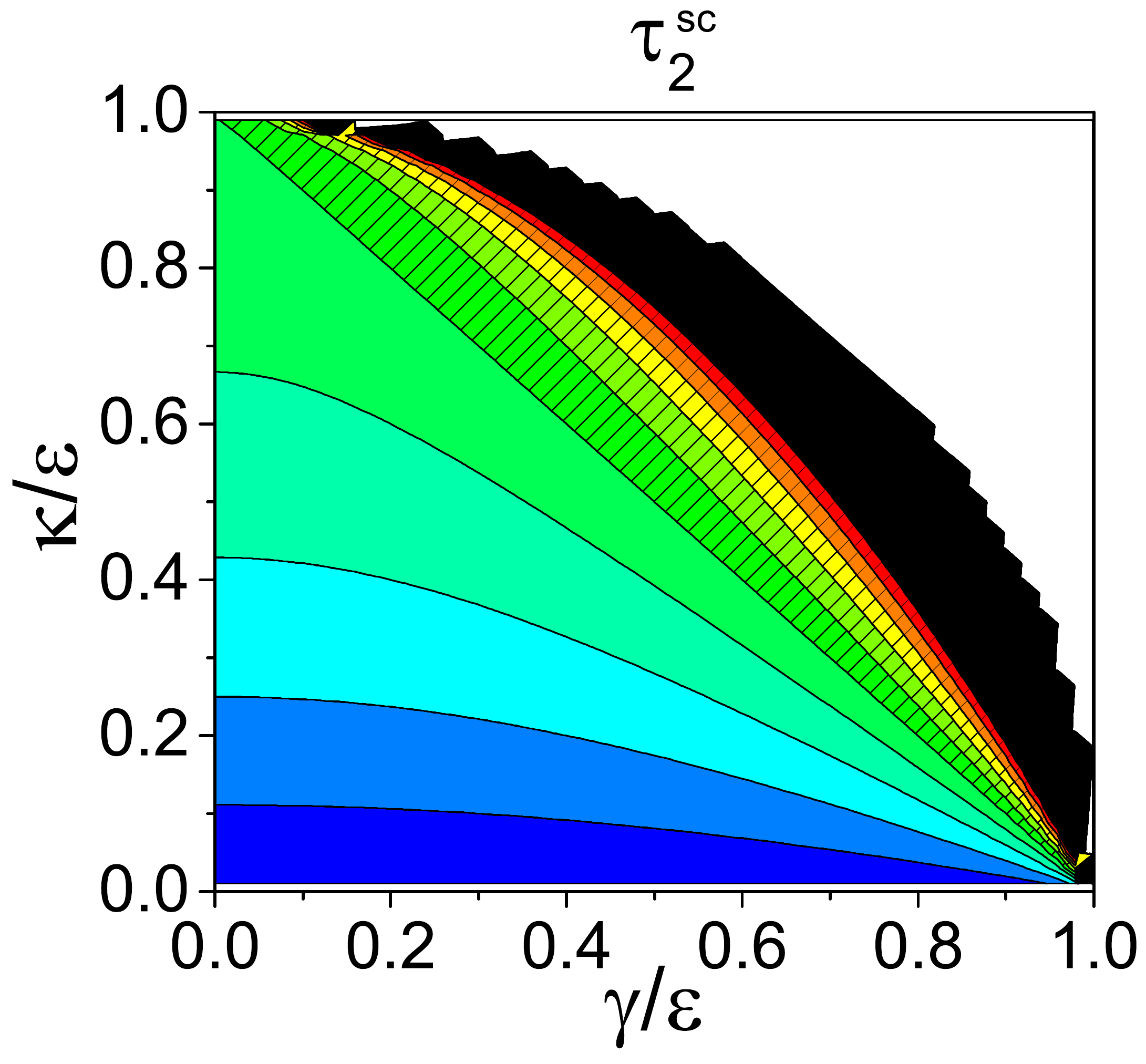}
 \includegraphics[width=0.26\hsize]{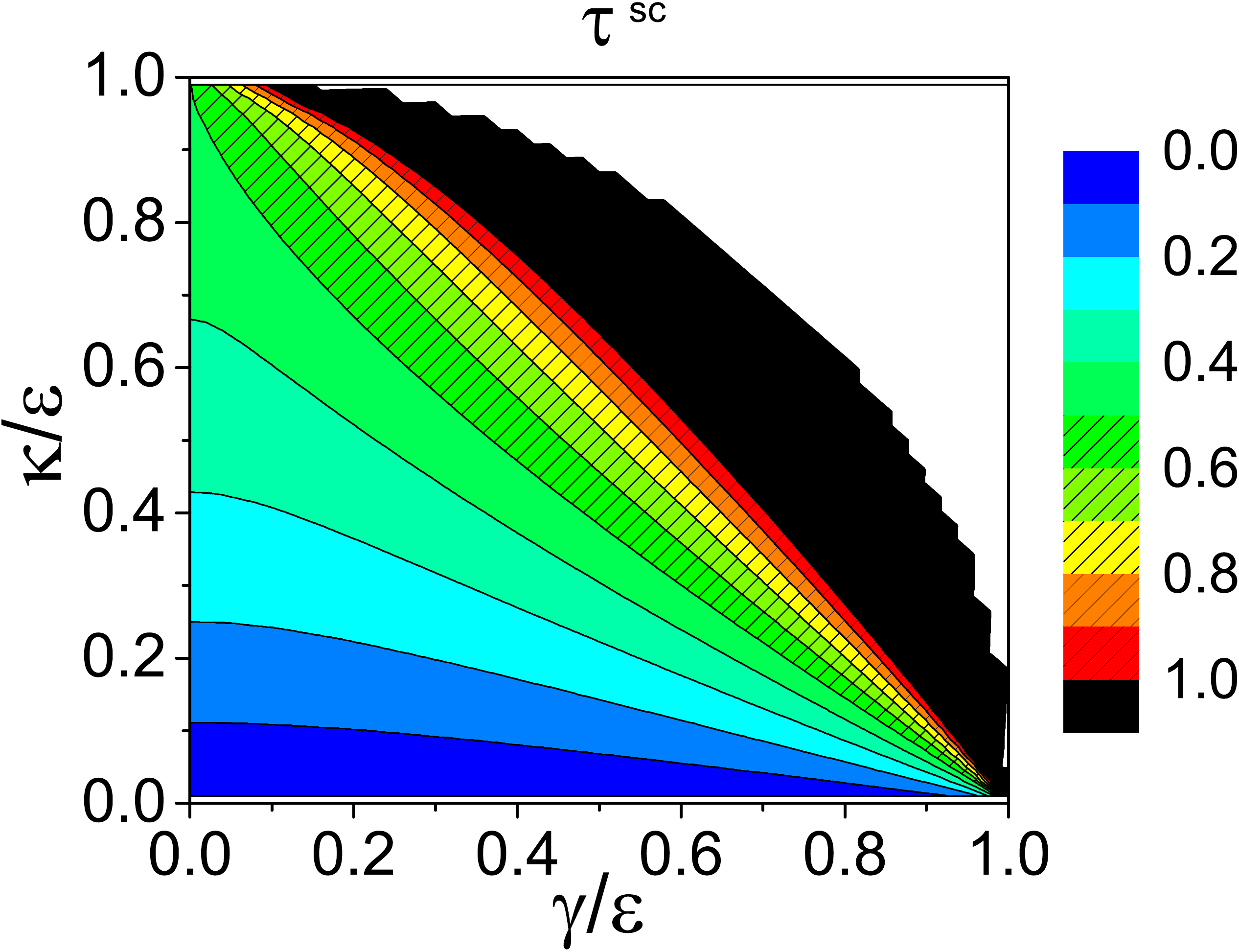}
 \includegraphics[width=0.26\hsize]{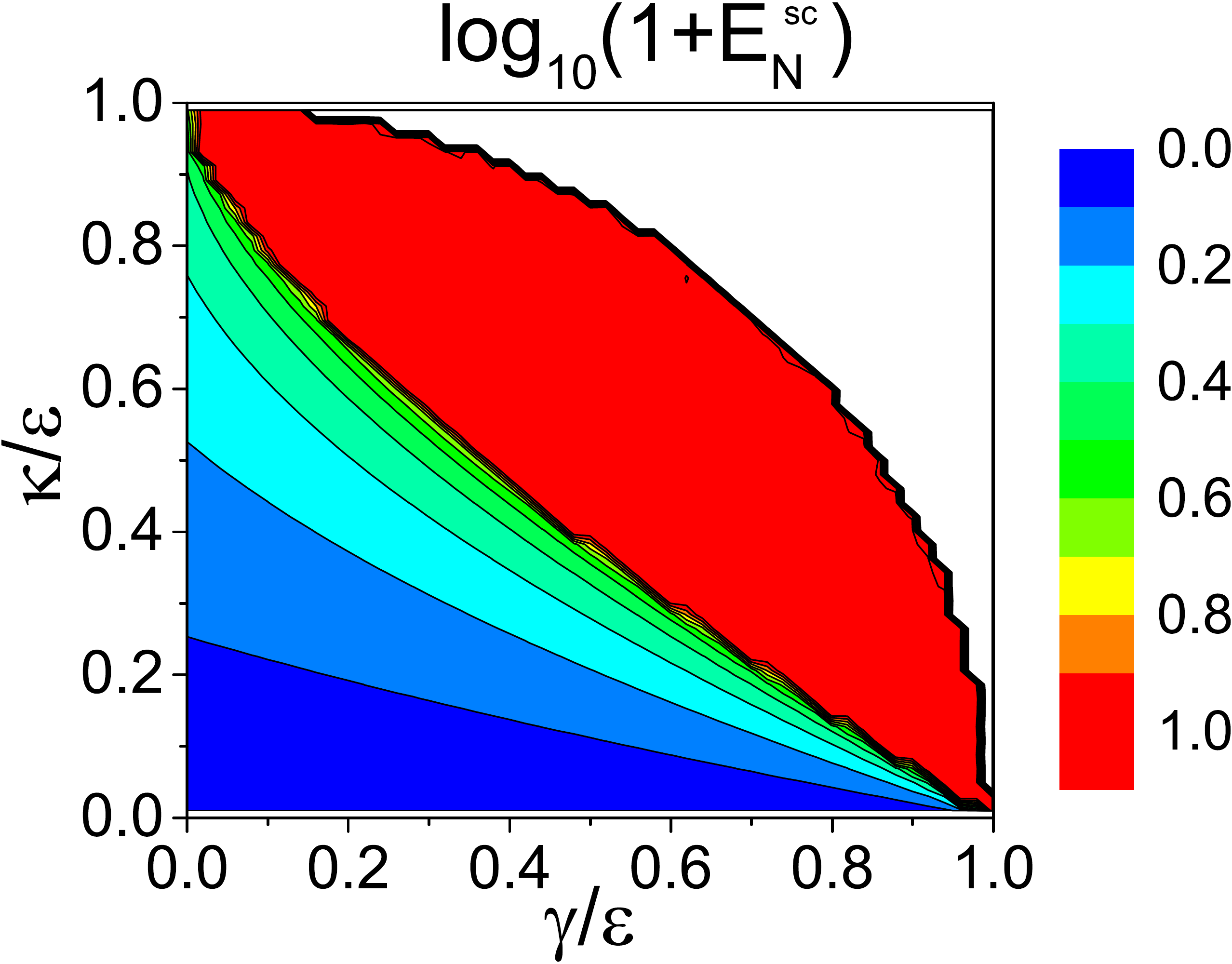}
 \vspace{2mm}
 \centerline{ \small (i) \hspace{.22\hsize} (j) \hspace{0.22\hsize} (k) \hspace{.22\hsize} (l)}

 \caption{Nonclassicality depths $ \tau_1 $ of mode 1 (a,e,i), $ \tau_2 $ of mode 2
 (b,f,j) and nonclassicality depths $ \tau $ (c,g,k) and negativities $ E_N
 $ (d,h,l) of the whole system versus dimensionless model parameters $ \kappa/\epsilon $ and $ \gamma/\epsilon $.
 Quantities with superscript id [sc] arise in the
 ideal (sink) model with partial reservoir fluctuations [semiclassical model with no reservoir fluctuations].
 In the graphs with the nonclassicality depths, nonphysical
 values $ \tau > 1 $ were determined in the black areas and
 values $ \tau > 0.5 $ not compatible with the Gaussian states were reached in the hatched colored areas.}
\label{fig2}
\end{figure*}

We compare the predictions of the ideal (sink) model with partial
reservoir fluctuations with those of the physically-consistent
model with complete reservoir fluctuations at a general level by
considering both nonclassicalities and entanglement. The maximal
values of nonclassicality depths of the whole system ($ \tau $)
and its constituting modes ($ \tau_1 $, $ \tau_2 $), as well as
the negativity $ E_N $ are drawn in Fig.~\ref{fig2} as they depend
on the system parameters.
The comparison of graphs of the nonclassicality depths plotted in
Figs.~\ref{fig2}(e,f,g) for the ideal (sink) model with those in
Figs.~\ref{fig2}(a,b,c) for the physically-consistent model
reveals that the nonclassicality depths $ \tau^{\rm id} $, $
\tau_1^{\rm id} $, and $ \tau_2^{\rm id} $ of the ideal (sink)
model are systematically greater than the nonclassicality depths $
\tau $, $ \tau_1 $, and $ \tau_2 $ of the physically-consistent
model with complete reservoir fluctuations. Moreover, whereas the
physically-consistent model gives correct values of the
nonclassicality depths even in the area of parameters with the
exponential behavior, the ideal (sink) model predicts the
nonclassicality depths only in the region of parameters with the
oscillatory behavior. Even in this region, in the area close to
the curve giving EPs, we find nonphysical values of the
nonclassicality depths $ \tau^{\rm id} $ greater than 1. The
regions in which the ideal (sink) model gives the values of the
nonclassicality depths $ \tau^{\rm id} \le 0.5 $ compatible with
the Gaussian form of the states are even smaller. The regions with
$ \tau^{\rm id} > 0.5 $ that contradict the Gaussian form of the
states are indicated by the hatched colored areas in
Figs.~\ref{fig2}(e,f,g). Similarly, the values of the negativity $
E_N^{\rm id} $ of the ideal (sink) model are plotted in
Fig.~\ref{fig2}(h). It is seen that they are systematically
greater than those of the physically-consistent model with
complete reservoir fluctuations in Fig.~\ref{fig2}(d). In general,
we may conclude that the ideal (sink) model, by partially
suppressing the reservoir fluctuations, systematically enhances
quantum features of the states in the system, as they manifest
themselves in the nonclassicality depths and negativity.

In Figs.~\ref{fig2}(i,j,k,l), we also plot the nonclassicality
depths $ \tau^{\rm sc} $, $ \tau_1^{\rm sc} $, and $ \tau_2^{\rm
sc} $ together with the negativity $ E_N^{\rm sc} $ of the
semiclassical model with no reservoir fluctuations. The analysis
of the behavior of this semiclassical model is very important, as
such models are frequently addressed in the literature. The reason
is that the omission of reservoir fluctuations allows to treat the
model at the level of the non-Hermitian Hamiltonian description,
which is considerably simpler than that based on the Liouvillian.
The comparison of graphs in Figs.~\ref{fig2}(i,j,k,l) with those
in Figs.~\ref{fig2}(a,b,c,d) determined for the
physically-consistent model with complete reservoir fluctuations
brings us to the conclusions similar to those made for the ideal
(sink) model: The semiclassical model without reservoir
fluctuations systematically overestimates both nonclassicality and
entanglement quantifiers. On the other hand, a detailed comparison
of the graphs in Figs.~\ref{fig2}(i,j,k) with those in
Figs.~\ref{fig2}(e,f,g) reveals that the areas of parameters where
the semiclassical model predicts the physically-acceptable values
of the nonclassicality depths are larger than those belonging to
the ideal (sink) model.

We note that, for experimental Gaussian fields, we may measure the
principle squeezing variance \cite{Luks1988} in homodyne detection
\cite{Leonhardt1997} or even simpler by using
photon-number-resolving detectors \cite{Barasinski2023} to infer
the values of the nonclassicality depth $ \tau $. The negativity $
E_N $ for a two mode field can then be conveniently obtained from
photon-number-resolved measurements using photon-number moments up
to fourth order \cite{Barasinski2023}.

\section{Comparison of the nonclassicality and entanglement evolution
in $ \mathcal{PT} $-symmetric systems with different levels of
reservoir fluctuations}

Differences observed in the extremal values of the nonclassicality
and entanglement quantifiers in the ideal (sink) model with
partial inclusion of reservoir fluctuations and the semiclassical
model with no reservoir fluctuations vs. the physically-consistent
model with complete reservoir fluctuations change/evolve in time.
We may identify two qualitatively different types of their
behavior. In the first one, the difference between the predictions
for a given quantifier gradually increases with time, i.e., the
relative difference develops from zero at the initial time. In
this case the predictions of the investigated models with
partial/no inclusion of reservoir fluctuations agree well with
those of the physically-consistent model for short times and, as a
rule of thumb, the longer is the time, the greater are the
differences. In the second type of behavior, a nonzero difference
of a given quantity occurs already at short times involving the
initial time. There is no prediction for the subsequent evolution
of this difference in this case and it may also decrease.

\begin{figure*}[t]  
 \includegraphics[width=0.225\hsize]{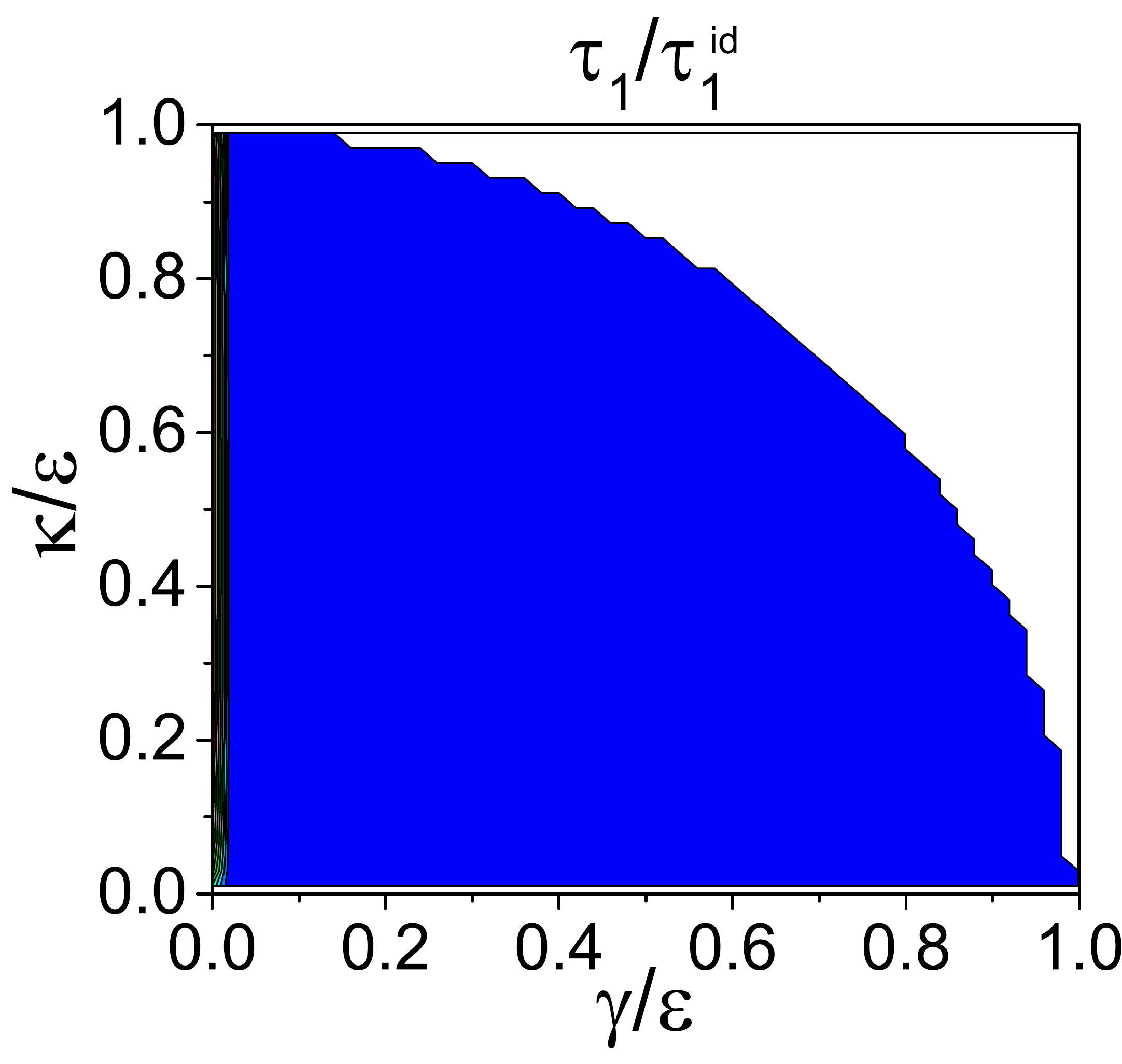}
 \includegraphics[width=0.225\hsize]{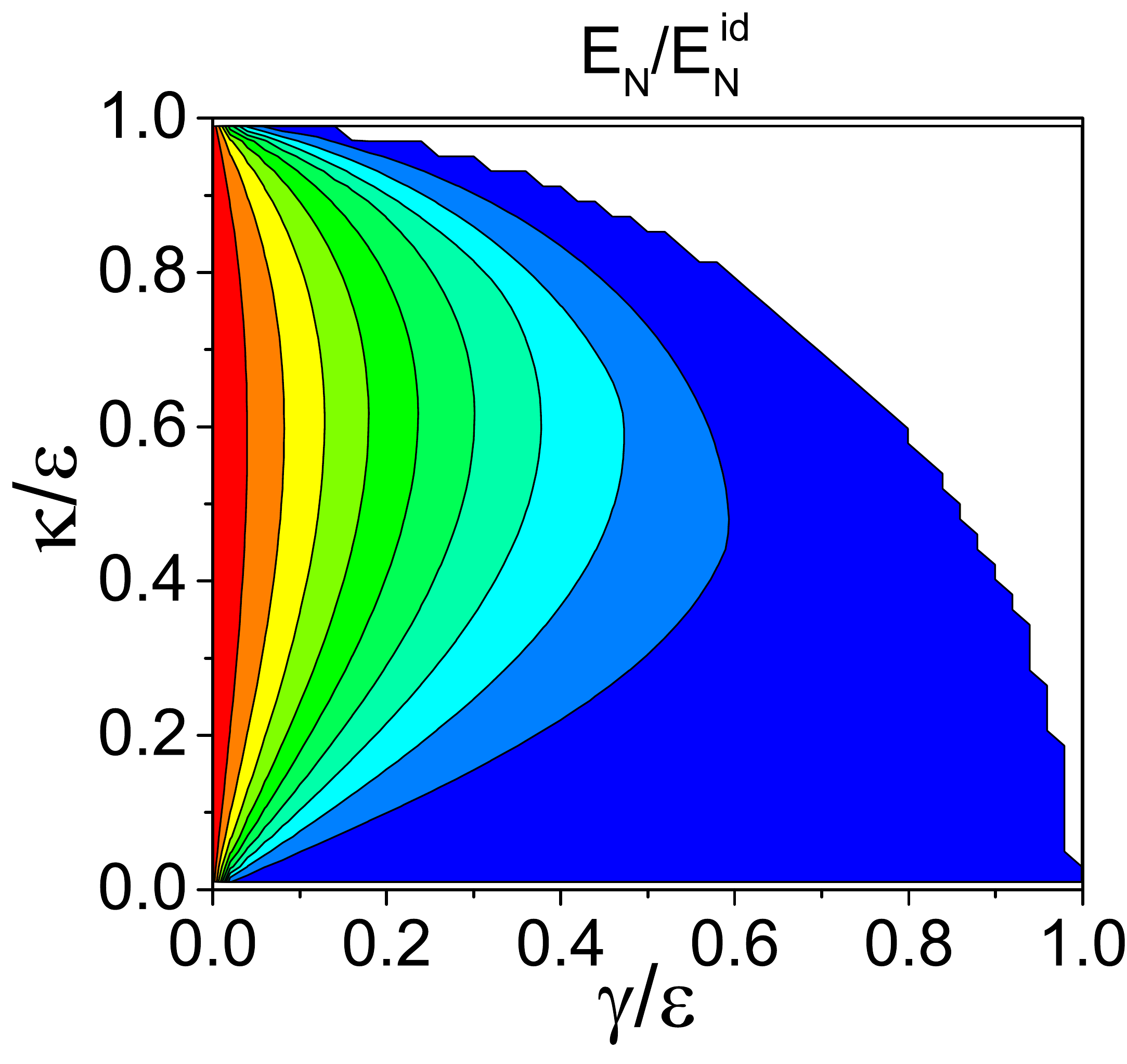}
 \includegraphics[width=0.225\hsize]{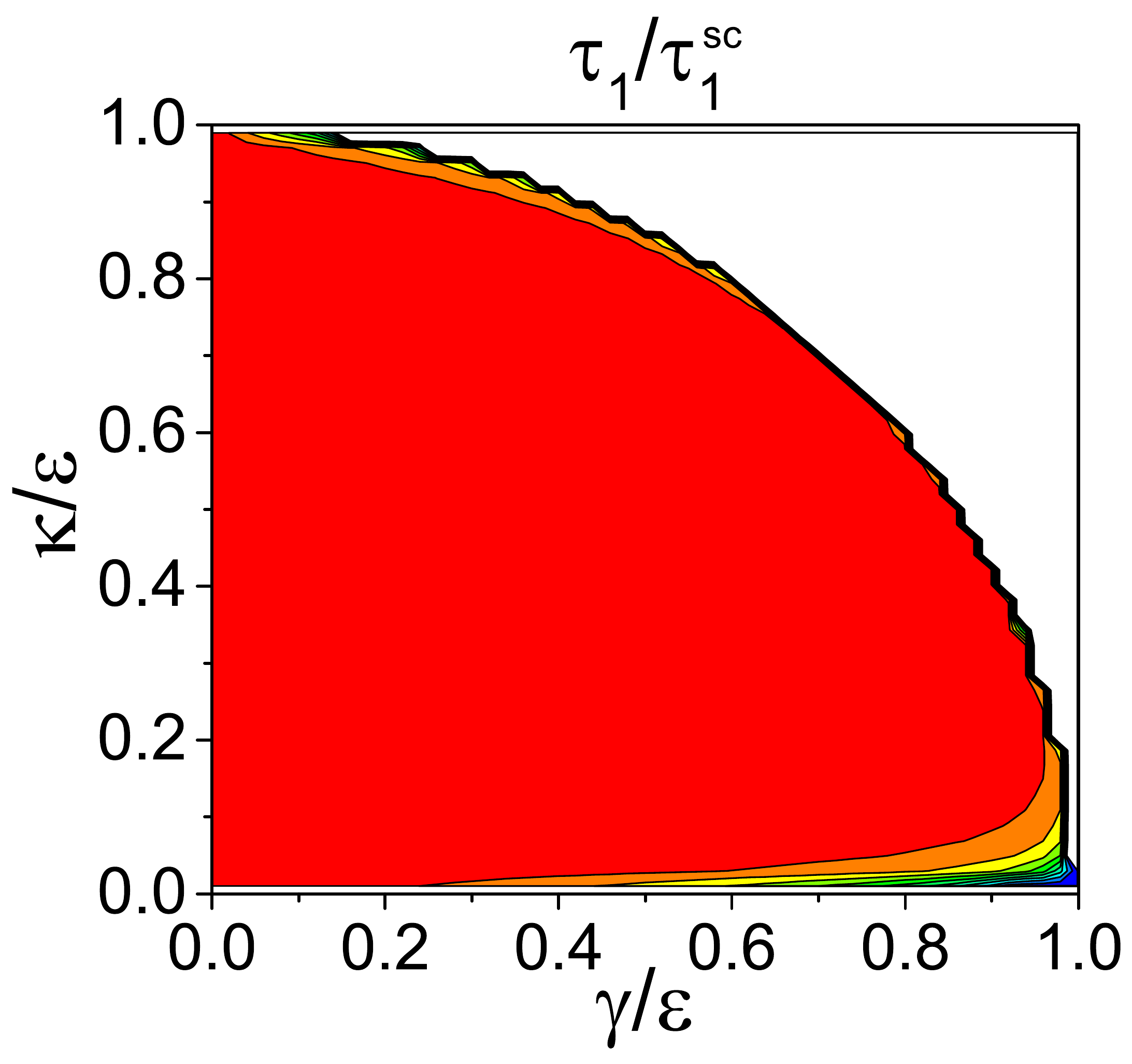}
 \includegraphics[width=0.26\hsize]{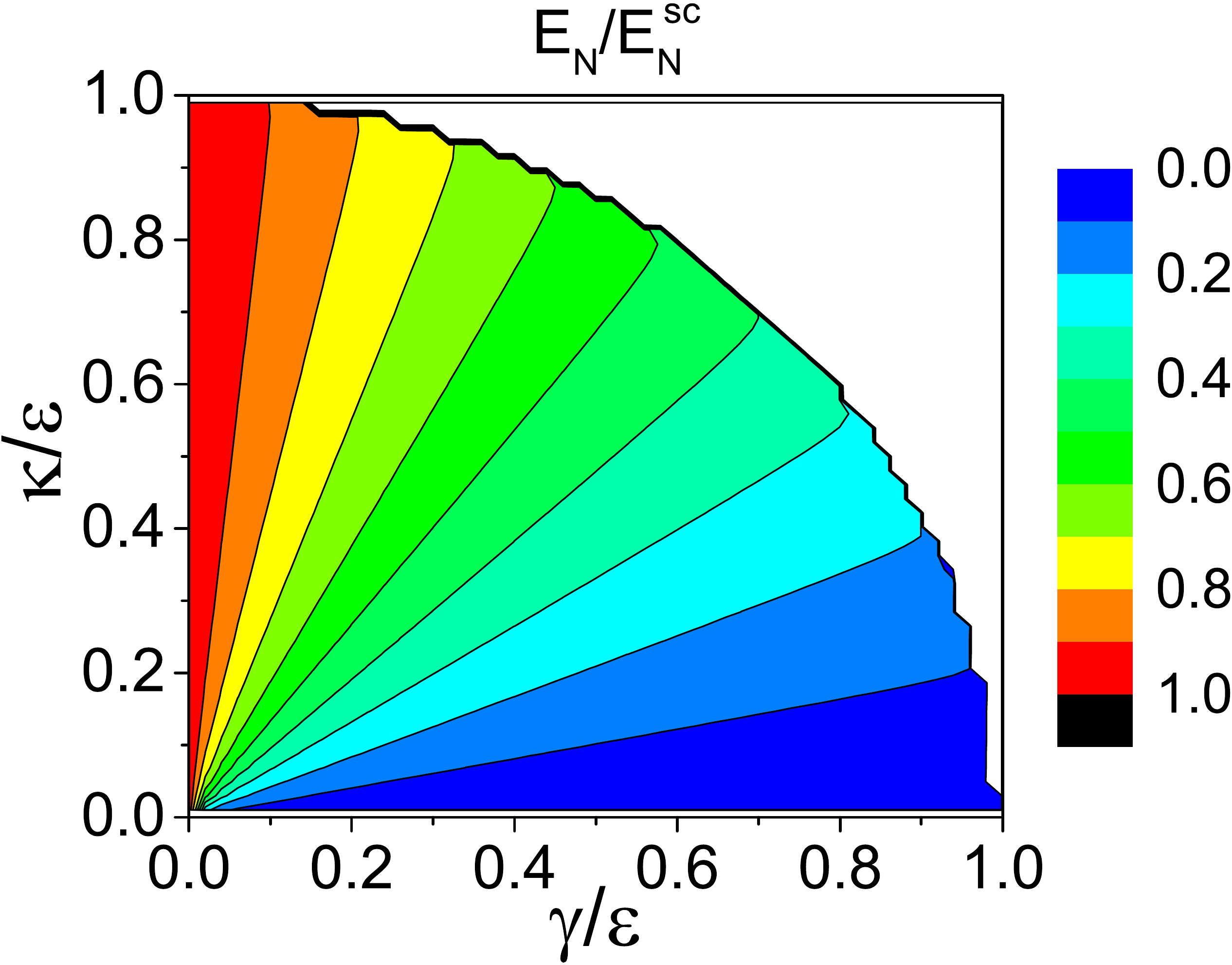}
 \vspace{2mm}
 \centerline{ \small (a) \hspace{.22\hsize} (b) \hspace{0.22\hsize} (c) \hspace{.22\hsize} (d)}

 \includegraphics[width=0.225\hsize]{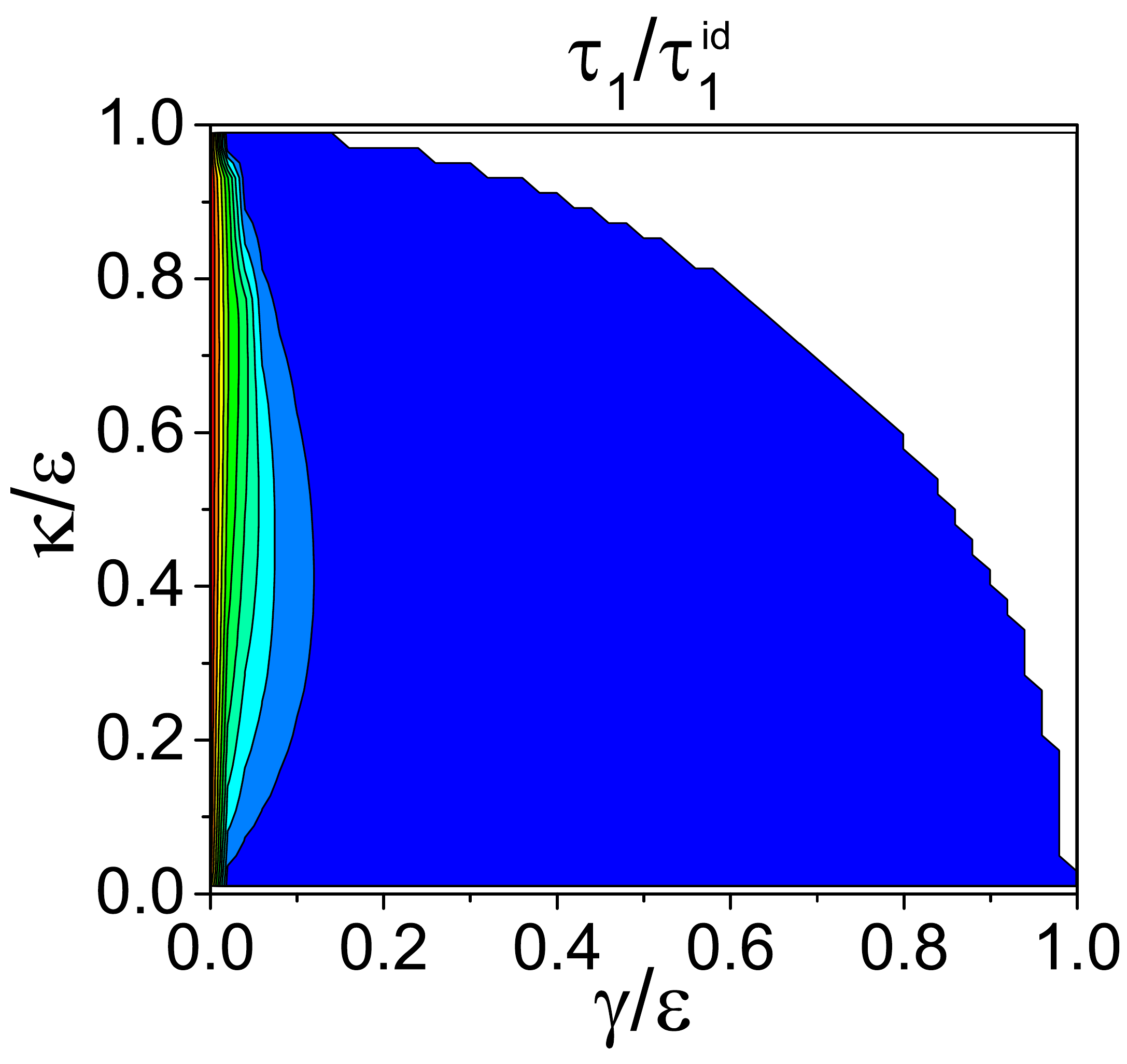}
 \includegraphics[width=0.225\hsize]{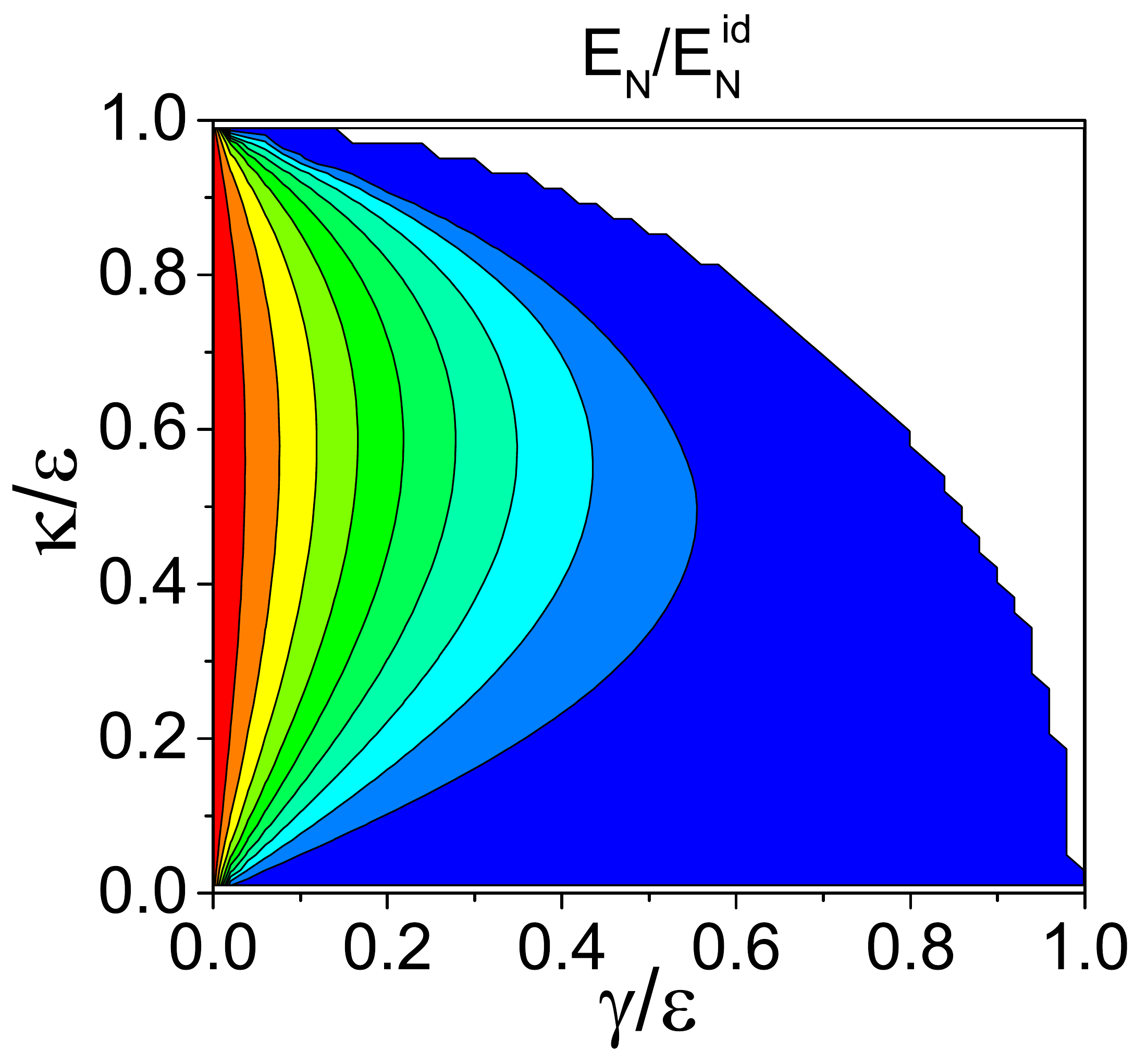}
 \includegraphics[width=0.225\hsize]{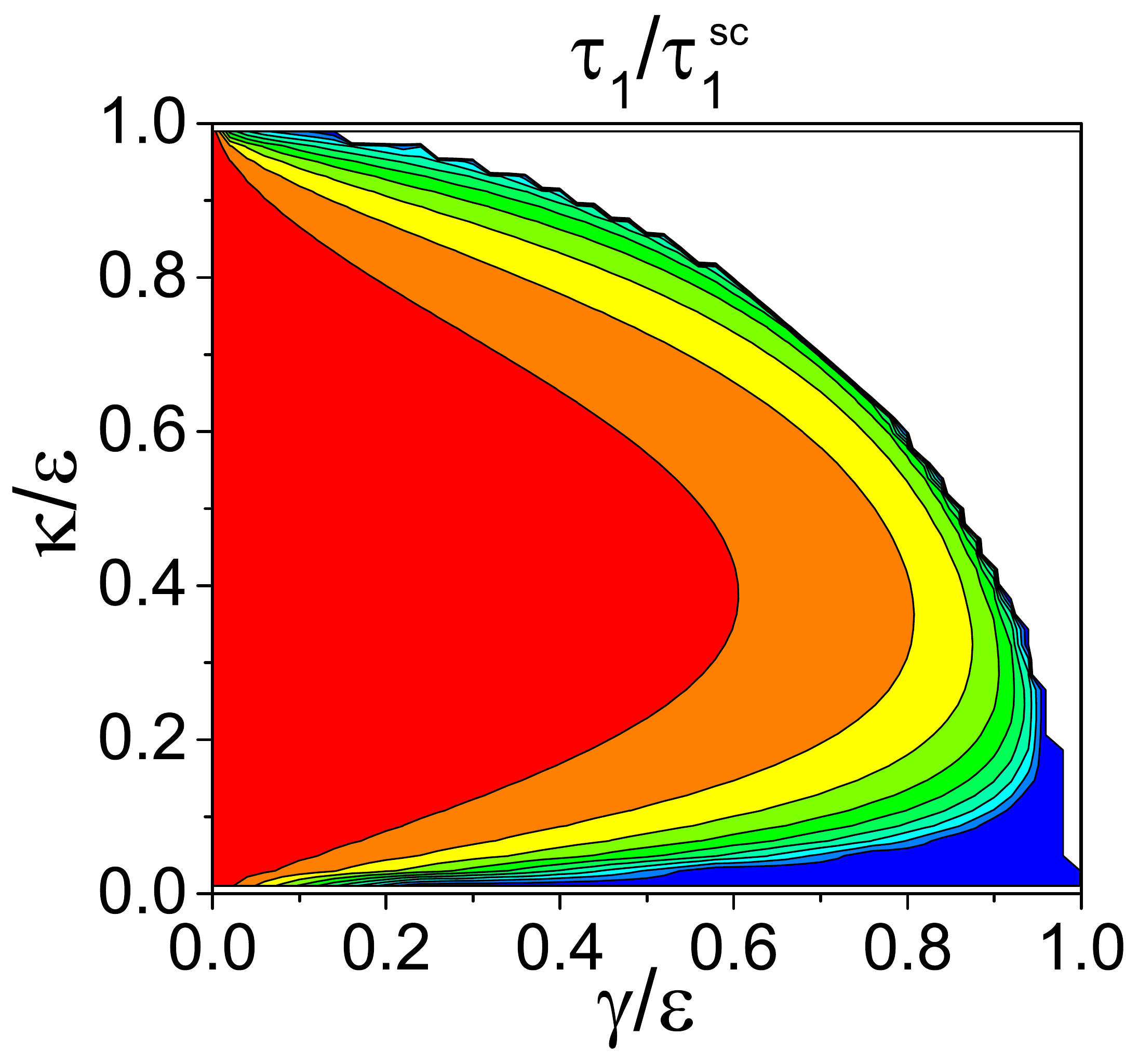}
 \includegraphics[width=0.26\hsize]{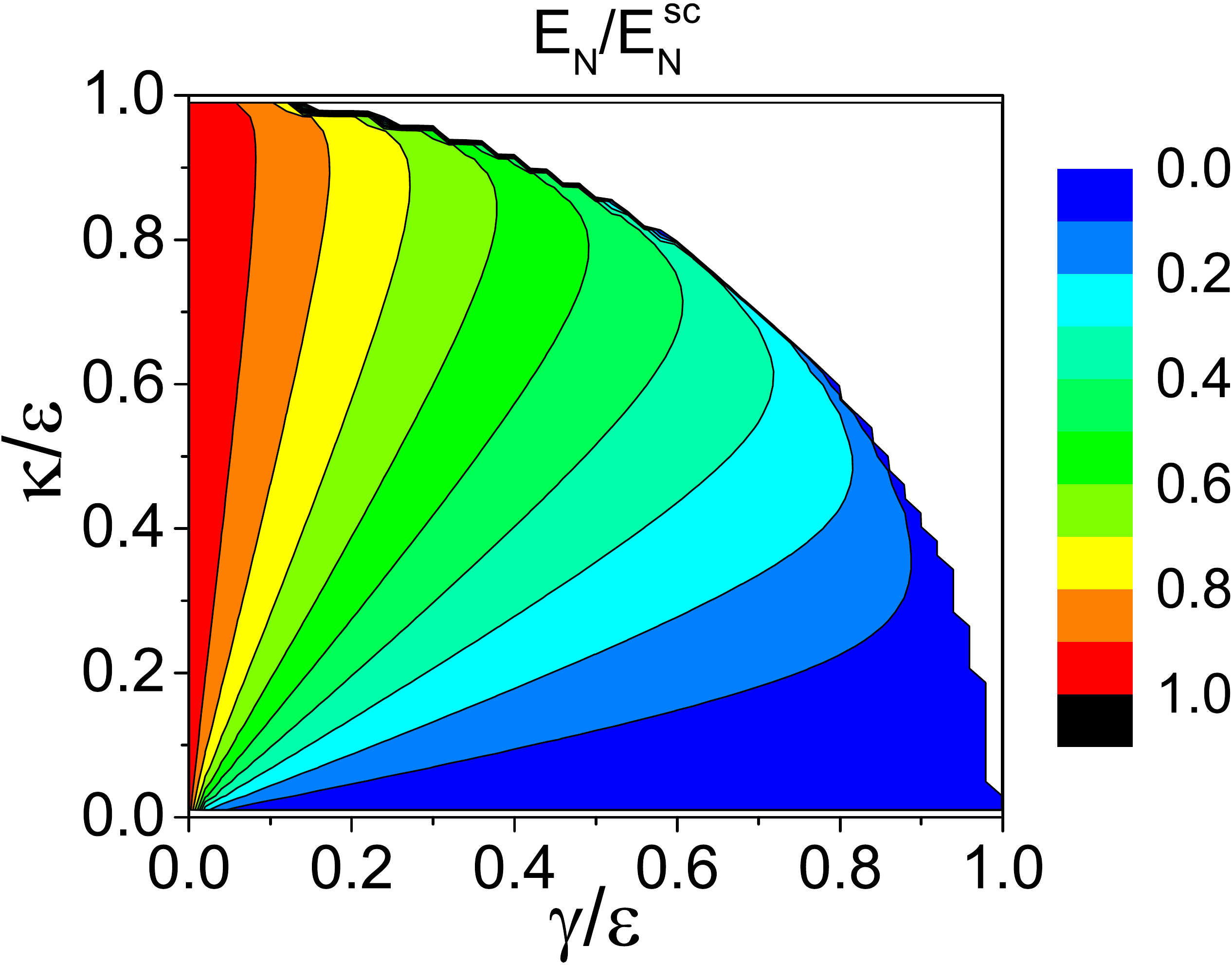}
 \vspace{2mm}
 \centerline{ \small (e) \hspace{.22\hsize} (f) \hspace{0.22\hsize} (g) \hspace{.22\hsize} (h)}

 \includegraphics[width=0.225\hsize]{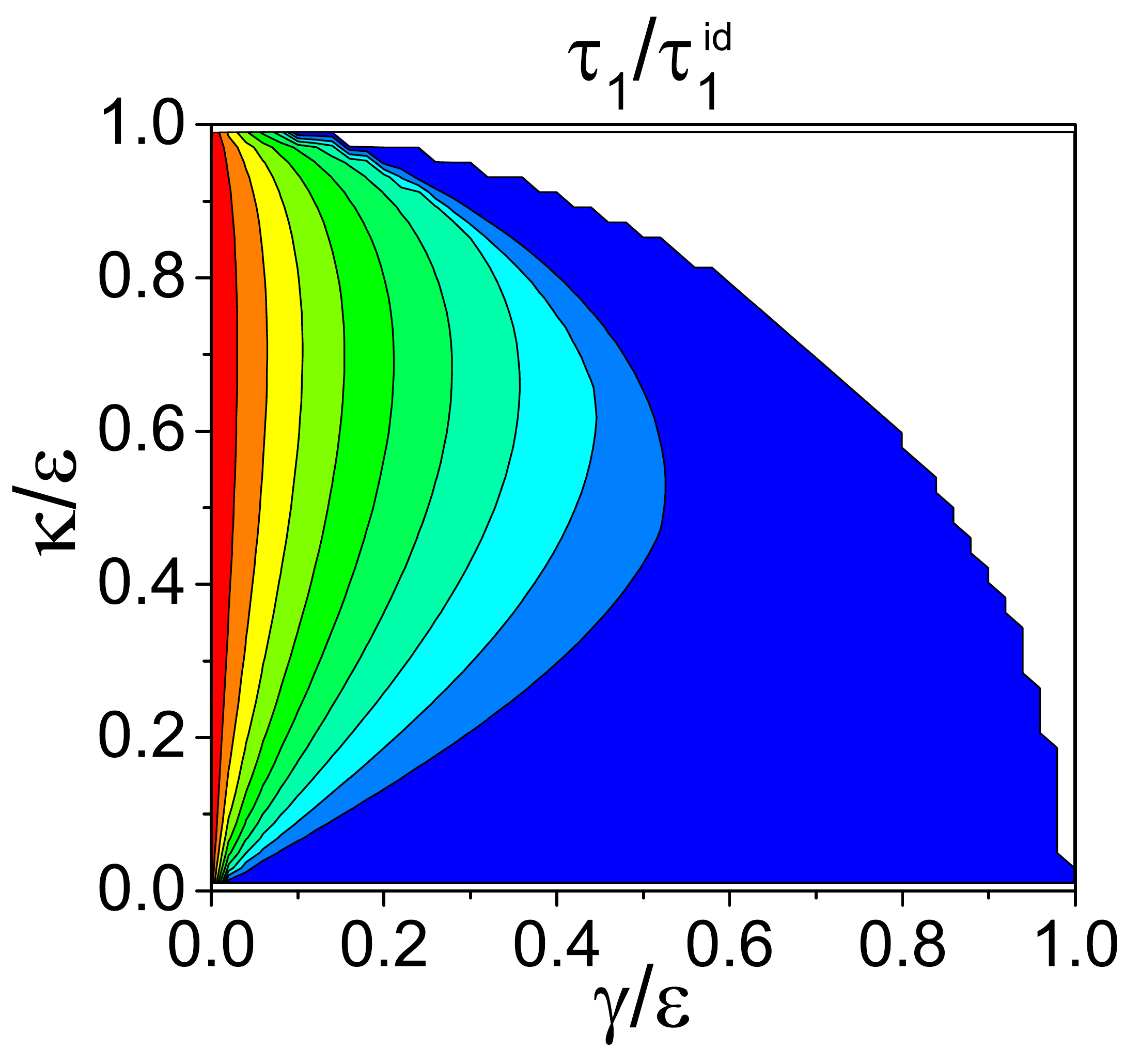}
 \includegraphics[width=0.225\hsize]{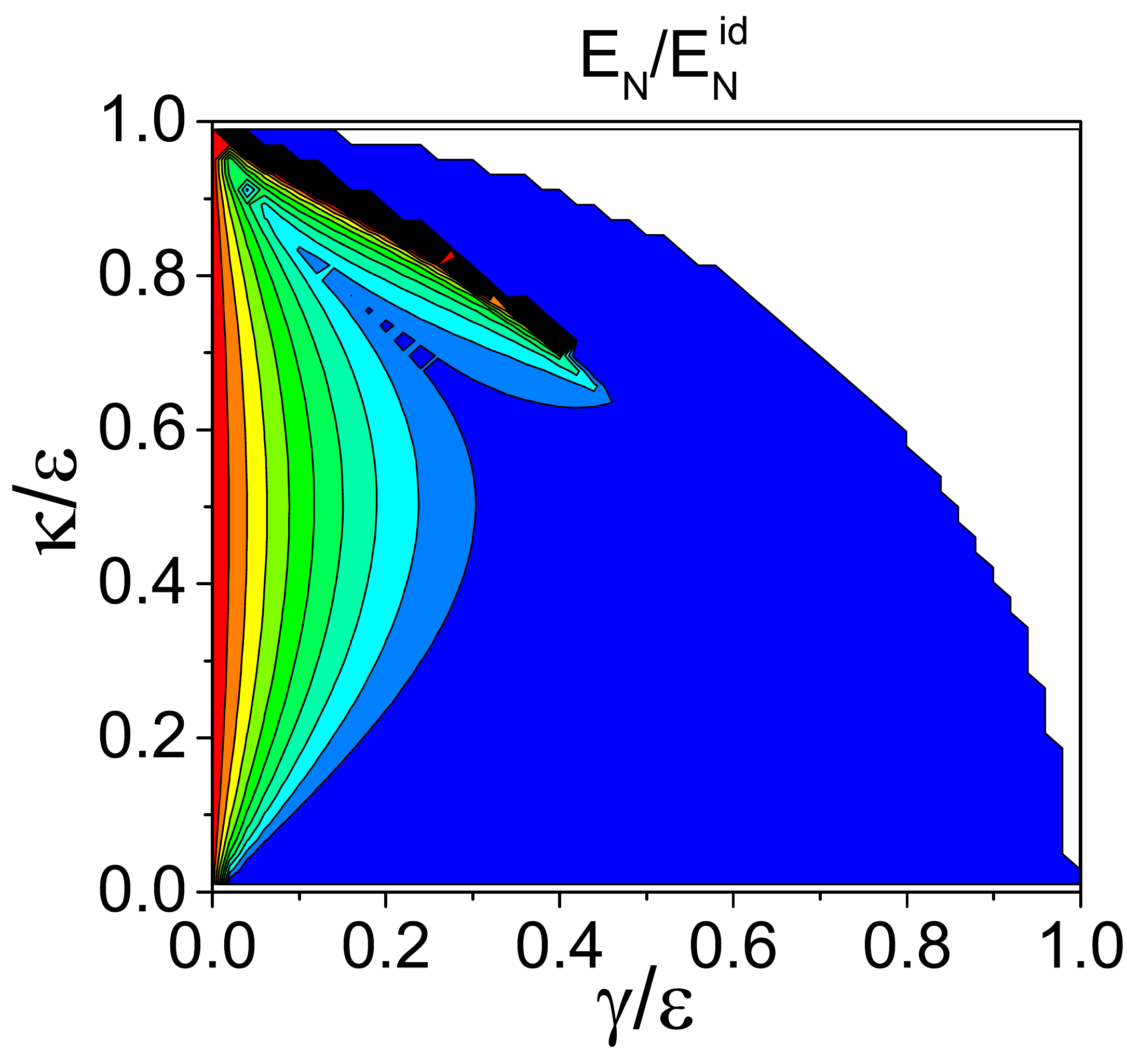}
 \includegraphics[width=0.225\hsize]{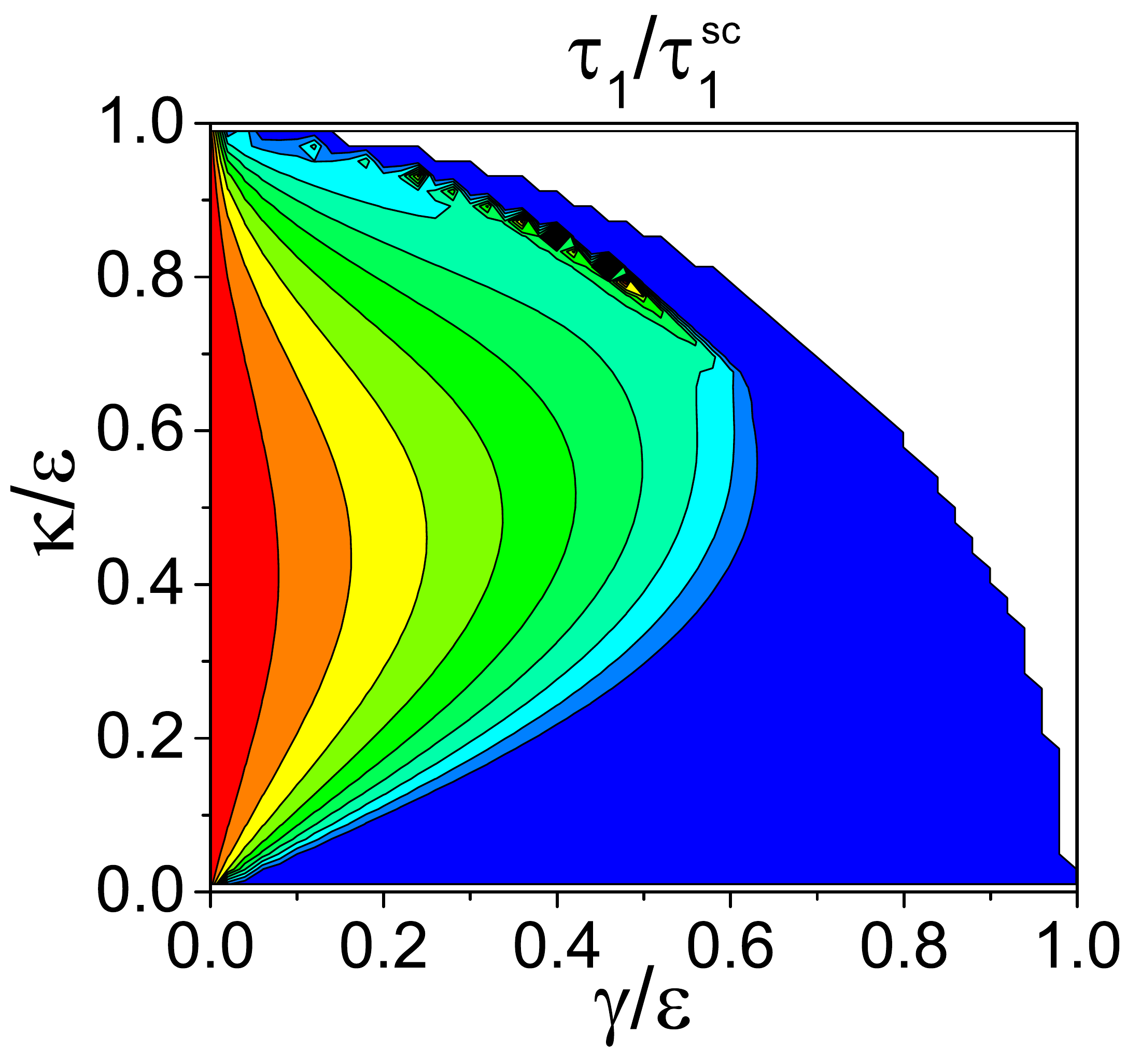}
 \includegraphics[width=0.26\hsize]{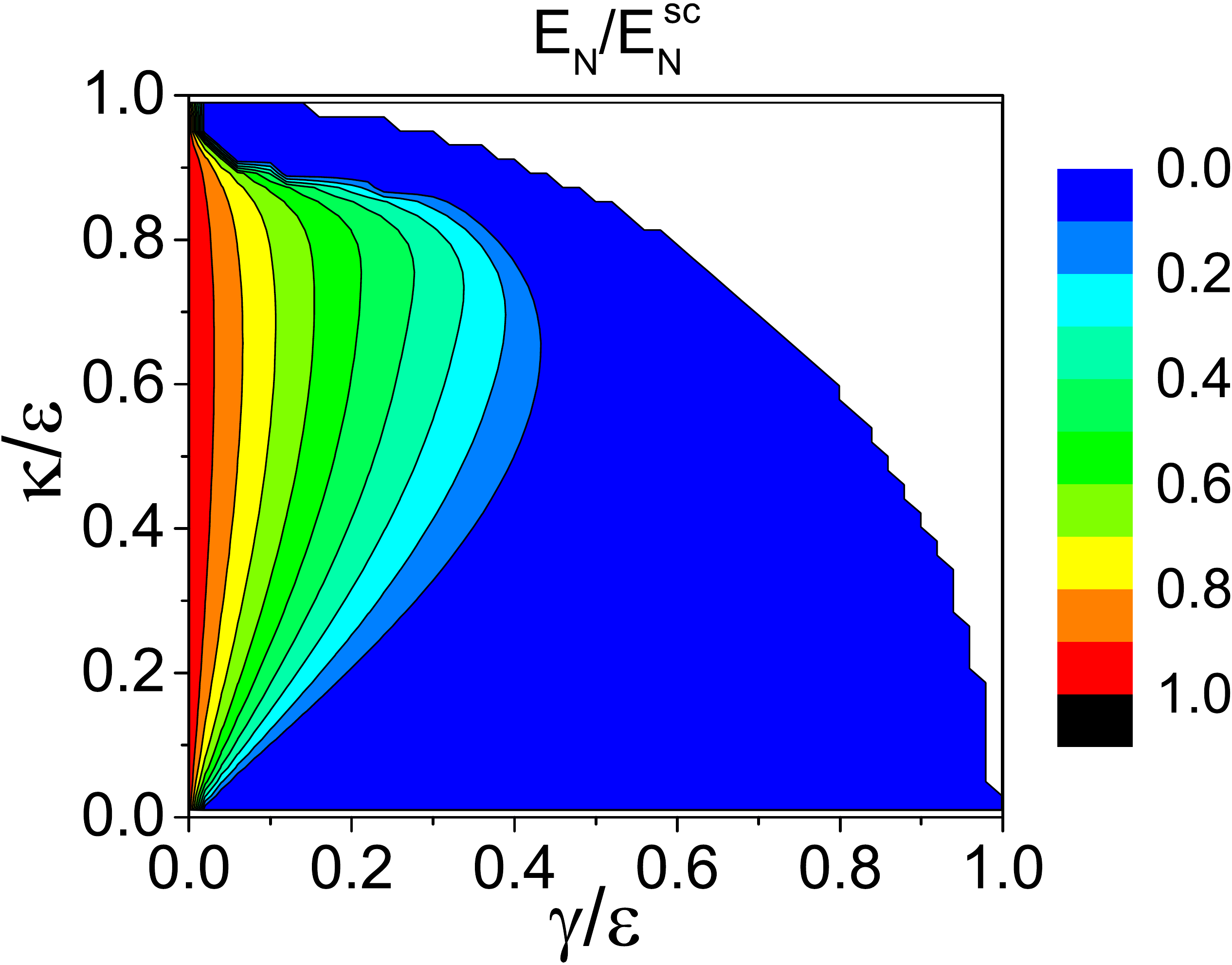}
 \vspace{2mm}
 \centerline{ \small (i) \hspace{.22\hsize} (j) \hspace{0.22\hsize} (k) \hspace{.22\hsize} (l)}

 \caption{Ratios of nonclassicality depths $ \tau_1/\tau_1^{\rm id} $ (a,e,i) and $ \tau_1/\tau_1^{\rm sc} $ (c,g,k) of mode 1
  and ratios of negativities $ E_N/E_N^{\rm id} $ (b,f,j) and $ E_N/E_N^{\rm sc} $ (d,h,l) of the whole system
  versus dimensionless model parameters
  $ \kappa/\epsilon $ and $ \gamma/\epsilon $. The ratios are determined for $ t_1= 10^{-3}T $ (a--d), $ t_2= 10^{-2}T $ (e--h) and
  $ t_3= 10^{-1}T $ (i--l) where $ T $ is given in Eq.~(\ref{40}).
 Quantities with superscript id [sc] arise in the
 ideal (sink) model with partial reservoir fluctuations [semiclassical model with no reservoir fluctuations].}
\label{fig3}
\end{figure*}

Both types of behavior are seen in Fig.~\ref{fig3}, in which we
plot the ratios of negativities $ E_N/ E_N^{\rm id} $ and $ E_N/
E_N^{\rm sc} $ and local nonclassicalities $ \tau_1/\tau_1^{\rm
id} $ and $ \tau_1/\tau_1^{\rm sc} $ of mode 1 in the
physically-consistent model vs. the ideal (sink) model and the
semiclassical model, respectively, in three subsequent time
instants $ t_1= 10^{-3}T $, $ t_2= 10^{-2}T $ and $ t_3= 10^{-1}T
$. These instants represent small fractions of the period $ T $,
\begin{equation} 
  T = 2\pi/\sqrt{ 1 - \frac{\kappa^2 + \gamma^2 }{ \epsilon^2}  } ,
\label{40}
\end{equation}
that characterizes the periodic behavior of the models with
partial/no inclusion of reservoir fluctuations.
As the ideal (sink) and semiclassical models have partially or
fully suppressed fluctuations, we expect greater values of the
nonclassicality depths $ \tau_1 $ and the negativities $ E_N $
arising in these models compared to those of the
physically-consistent model. Indeed, the ratios plotted in
Fig.~\ref{fig3} are smaller or equal to one for the vast majority
of the system parameters. We note that the predictions of all the
three models coincide for $ \gamma/\epsilon = 0 $. Whereas the
ratios of the negativities $ E_N/ E_N^{\rm id} $ and $ E_N/
E_N^{\rm sc} $ of both models with partial/no reservoir
fluctuations and ratios of the nonclassicality depths $
\tau_1/\tau_1^{\rm sc} $ of the semiclassical model start from 1
in the limit $ t \rightarrow 0 $, the ratios of the
nonclassicality depths $ \tau_1/\tau_1^{\rm id} $ of the ideal
(sink) model are nonunit for very short times $ t $. Comparing the
graphs in Fig.~\ref{fig3} for time instants $ t_1 $, $ t_2 $, and
$ t_3 $ (plotted in different rows), we observe the decrease of
the ratios $ E_N/ E_N^{\rm id} $, $ E_N/ E_N^{\rm sc} $, and $
\tau_1/\tau_1^{\rm sc} $ as the time increases. On the other hand,
the ratio $ \tau_1/\tau_1^{\rm id} $ being very small for very
short times increases with time. We note that the
physically-consistent model does not behave periodically owing to
the constantly and irreversibly acting reservoir fluctuating
forces. Nevertheless, the values of the nonclassicality depths $
\tau_1 $ and the negativities $ E_N $ reached in the interval $
\langle 0, T\rangle $, i.e., in the first period of the models
with partial/no reservoir fluctuations, are usually greater than
those obtained for longer times.

According to the fluctuation-dissipation theorem
\cite{Meystre2007}, the strength of the second-order correlation
functions of the reservoir fluctuating forces depends linearly on
the damping/amplification parameter $ \gamma $. Thus, the greater
is the ratio $ \gamma/\epsilon $, (i) the stronger are the
fluctuating forces, (ii) the more detrimental are the effects on
the nonclassicality and entanglement, and (iii) the smaller are
the ratios of the nonclassicality depths and negativities. This
consideration valid for shorter times is documented in the graphs
of Fig.~\ref{fig3}. The ratio of the negativities $ E_N/ E_N^{\rm
id} $ drawn in Fig.~\ref{fig3}(j) for $ t_3= 10^{-1}T $
demonstrates departure from this rule valid for shorter times.

Fixing the ratio $ \gamma/\epsilon $ and assuming again shorter
times, the ratios of the negativities and nonclassicality depths
in Fig.~\ref{fig3} attain maximum in the interval $
\kappa/\epsilon \in \langle 0, \sqrt{1 - \gamma^2/\epsilon^2}
\rangle $. The increase of these ratios with the increasing $
\kappa/\epsilon $ on the left-hand side of this maximum is
attributed to the increase of the system nonlinearity. Greater
values of the relative nonlinearity $ \kappa/\epsilon $ mean
faster nonclassicality and entanglement generation that prevails
the detrimental role of the reservoir fluctuations. On the other
hand, the decrease of these ratios on the right-hand-side of the
maximum is attributed to the increased period $ T $ (and, thus,
the increased time instants $ t_{1,2,3} $) with the increasing
ratio $ \kappa/\epsilon $ that brings longer action of the
reservoir fluctuating forces.

The ratios of the negativities and nonclassicality depths are
rather small in the area close to the curve for the EPs. This is a
consequence of the fact that the period $ T $ of the models with
partial/no reservoir fluctuations is very long in this area and so
the detrimental effect of reservoir fluctuations is strong. We
note that the period $ T $ goes to infinity at the EPs which
results in the ratios determined at asymptotically long times.

These results obtained for specific time instants show, similarly
as the results for maximal values of the nonclassicality depths
and negativities in Sec.~IV, that the applicability of the ideal
(sink) model and the semiclassical model with partial/no inclusion
of reservoir fluctuations for our predictions of the nonclassical
properties of the studied $ {\cal PT} $-symmetric system is rather
limited. Whereas the semiclassical model gives reliable
predictions for shorter times, some of the predictions of the
ideal (sink) model may even be misleading at short times.

\section{Conclusions}

An analytical solution of the quantum-consistent
$\mathcal{PT}$-symmetric model of two nonlinearly interacting
damped and amplified bosonic modes coupled to reservoirs has been
obtained. Using this solution the evolution of the nonclassicality
and the entanglement of the generated Gaussian states has been
analyzed in the whole space of model parameters. Whereas both
nonclassical and entangled states are generated for shorter times,
the reservoir fluctuations suppress their generation for long
times. The analytical solution has allowed us to identify the
reservoir noise contribution that increases linearly with time and
causes a gradual loss of the nonclassicality and entanglement for
long times.

To understand the origin of this degradation of the system ability
to generate nonclassical and entangled states, we have considered
two simplified models with a partial and full suppression of
reservoir fluctuations: (1) a semiclassical model with no
reservoir fluctuations and (2) an ideal (sink) model with a
partial inclusion of reservoir fluctuations. Both models provide
periodic solutions which allow for the generation of nonclassical
and entangled states even for long times. The models differ by the
level of their physical consistency: Whereas the semiclassical
model violates the fluctuation-dissipation theorem, the ideal
(sink) model obeys specific fluctuation-dissipation relations.
However, its modified reservoir is endowed with the properties of
the sink which removes the noise from the system.

Because of the partial or full suppression of reservoir
fluctuations, both models provide systematically greater values of
the nonclassicality depths as measures of quantumness and the
negativity as a measure of entanglement. Unfortunately, the
attained values of the nonclassicality depths may even exceed
their physically allowed ranges, in the area of the model
parameters around the EPs. Whereas the semiclassical model gives
reliable predictions for short times, the ideal (sink) model,
though being more physically consistent, may provide misleading
results even for short times.

None of these two models, that allow for the generation of
nonclassical and entangled states in quantum
$\mathcal{PT}$-symmetric systems, can be applied to predict the
system behavior for longer times. The only physically-consistent
model is provided by the statistical physics of open quantum
systems (using the Liouvillians or alternatively the
Heisenberg-Langevin equations). This model properly includes the
reservoir fluctuations associated with the damping and
amplification of the system and, thus, describes its evolution in
a physically consistent way. This model, however, shows that the
irreversible reservoir fluctuations inevitably degrade the
nonclassicality and entanglement generated in the system, which
results in their complete loss for longer times. Thus, we find no
possibility for mutual compensation of the reservoir fluctuations
associated with the damping and amplification in the model. This
limitation qualitatively differs from a direct action of the
damping and amplification in the evolution of quantum
$\mathcal{PT}$-symmetric systems, in which they mutually interfere
to give a periodic evolution.

These results bring us to the general conclusion that the
detrimental role of reservoir fluctuations in quantum
$\mathcal{PT}$ -symmetric systems with damping and amplification
cannot be avoided and the suppression of nonclassicality and
entanglement in their evolution is their natural property that has
to be accepted.


\begin{acknowledgments}
J.P. thanks Anton\' \i n Luk\v{s} for discussions and reading the
manuscript. A.M. was supported by the Polish National Science
Centre (NCN) under the Maestro Grant No. DEC-2019/34/A/ST2/00081.
J.K.K. and W.L. are grateful for the support of the Polish
Minister of Education and Science given under the program
''Regional Initiative of Excellence'' in 2019-2023, project No.
003/RID/2018/19, funding amount PLN 11 936 596.10.
\end{acknowledgments}


%

\end{document}